\newcommand{\bbar}{\ensuremath{\mathchar'26\mkern-9mu b\mkern2mu} }

\documentclass[12pt,twoside,letterpaper,twocolumn]{article}
\pdfoutput=1
\usepackage[margin=20mm]{geometry}

\usepackage{url}
\usepackage{graphicx}
\usepackage{amssymb,amsmath}

\begin{document}
\author{Robert Brady and Ross Anderson\\
        \small University of Cambridge Computer Laboratory\\
        \small JJ Thomson Avenue, Cambridge CB3 0FD, United Kingdom\\
        \small \texttt{\{robert.brady,ross.anderson\}@cl.cam.ac.uk}}
\title{Why bouncing droplets are a pretty good model\\of quantum mechanics}

\date{\today}

\twocolumn[
  \begin{@twocolumnfalse}
    \maketitle
\begin{abstract}
  In 2005, Couder, Proti\`ere, Fort and Badouad showed that oil droplets 
  bouncing on a vibrating tray of oil can display nonlocal interactions  
  reminiscent of the particle-wave associations in quantum mechanics;
  in particular they can move, attract, repel and orbit each other. 
  Subsequent experimental work by Couder, Fort, Proti\`ere, Eddi,
  Sultan, Moukhtar, Rossi, Mol\'a\v{c}ek, Bush and Sbitnev has established
  that bouncing drops exhibit single-slit and double-slit diffraction,
  tunnelling, quantised energy levels, Anderson localisation and the
  creation/annihilation of droplet/bubble pairs.

  In this paper we explain why. We show first that the surface waves guiding
  the droplets are Lorentz covariant with the
  characteristic speed $c$ of the surface waves; 
  second, that pairs of bouncing droplets experience
  an inverse-square force of attraction or repulsion 
  according to their relative phase,
  and an analogue of the magnetic force;  third, that bouncing droplets
  are governed by an analogue of Schr\"odinger's equation where Planck's
  constant is replaced by an appropriate constant of the motion; and fourth,
  that orbiting droplet pairs exhibit spin-half symmetry
  and align antisymmetrically as in the Pauli exclusion principle. 
  Our analysis explains the similarities
  between bouncing-droplet experiments and the behaviour of quantum-mechanical
  particles. It also enables us to highlight some differences, and to predict
  some surprising phenomena that can be tested in feasible experiments.
\end{abstract}
  \end{@twocolumnfalse}
  ]


\section{Introduction}
\label{sec:introduction}

In 1978 Walker reported that a droplet of soapy water could bounce for several minutes
on a vibrating dish of the same fluid~\cite{walker1978drops}. In 2005 Couder, 
Proti\`ere, Fort and Badouad started the systematic study of this phenomenon
using droplets of silicone oil; the droplets can be made to bounce indefinitely on an oil 
surface that is vibrated vertically, and with the right amplitude and frequency
of vibration, droplets can move laterally or `walk'~\cite{couder2005dynamical}.
A thin film of air between the droplet and the surface prevents coalescence. 

\begin{figure}[htb]
	\centering
		\includegraphics[width=0.35\textwidth]{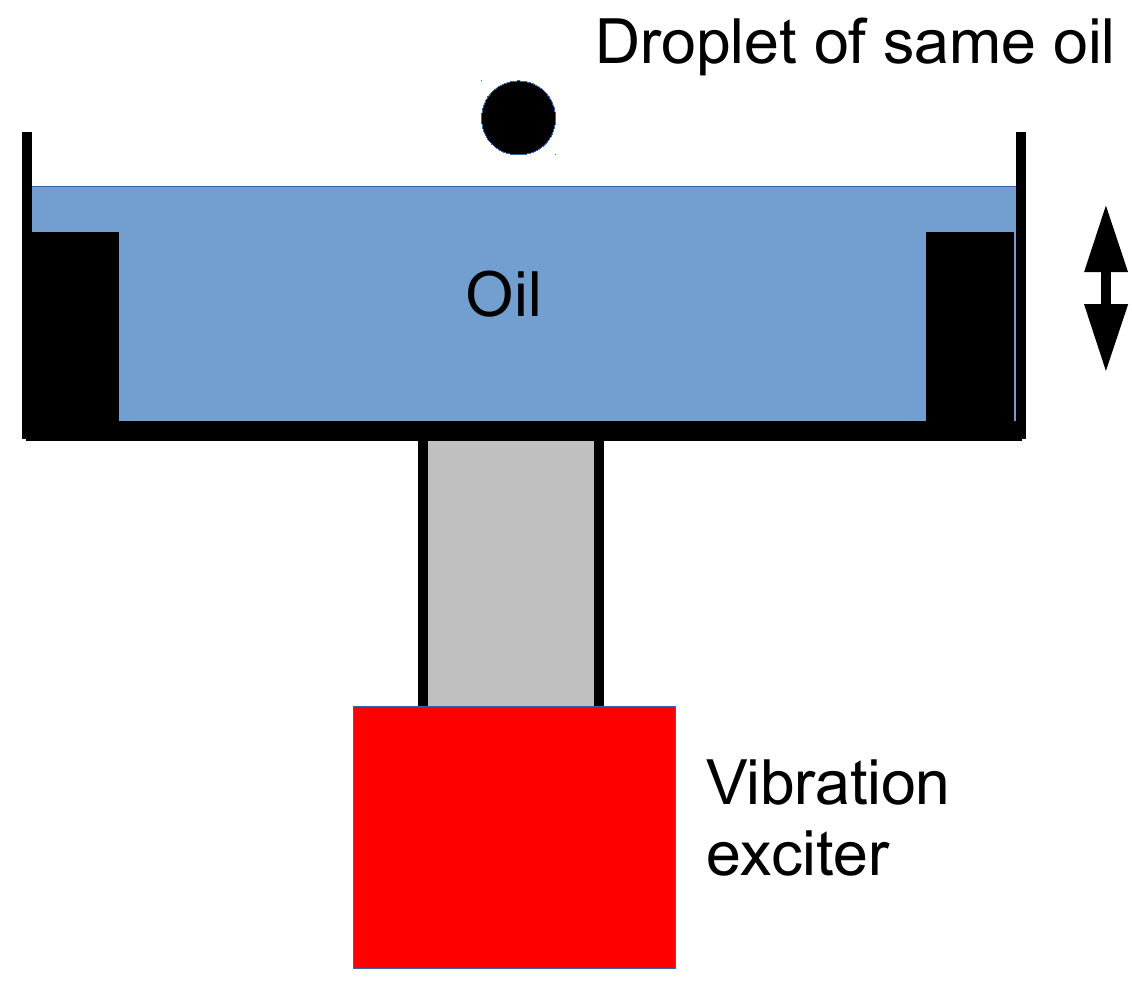}
	\caption{\em The experimental apparatus. The shape of the container
	reduces unwanted waves from the edge.}
	\label{fig:couder-experiment}
\end{figure}

Figure \ref{fig:couder-experiment} 
illustrates the apparatus, figure \ref{fig:couder2006prlbounce} has six
photographs of a bouncing droplet, and figure \ref{fig:Bouncing-feedback}
illustrates the vertical motion as a function of time. Here the droplet touches
down every other cycle; at lower driving amplitudes there are less interesting modes of
bouncing, where the droplet grazes off the peak near $f$ in the figure or
touches down every cycle. 

These experiments have since been reproduced in
laboratories around the world, such as by Mol\'a\v{c}ek and
Bush~\cite{molacek2013dropsbouncing}, and have appeared on
TV~\cite{couder2012tv}. They are of interest because the droplets exhibit much
of the behaviour that had previously been thought unique to quantum-mechanical
particles.

\section{Basic mechanism}
\label{sec:basicmechanism}
\label{sec:surface-waves}

\begin{figure}[ht]
	\centering
	\includegraphics[width=0.45\textwidth]{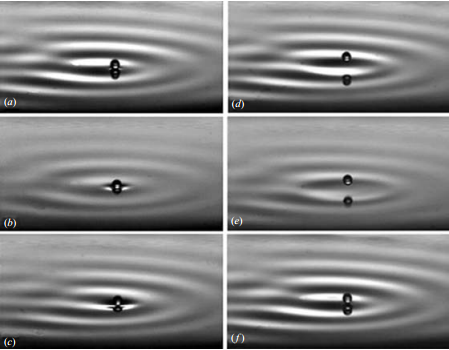}
	\caption{\em A droplet of silicone oil bouncing on the surface of the same liquid 
	which is vibrated vertically. 
	(courtesy Suzie Proti\`ere, Arezki Boudaoud and Yves Couder \cite{protiere2006particle})}
	\label{fig:couder2006prlbounce}
\end{figure}

\begin{figure}[htb]
	\centering
		\includegraphics[width=0.50\textwidth]{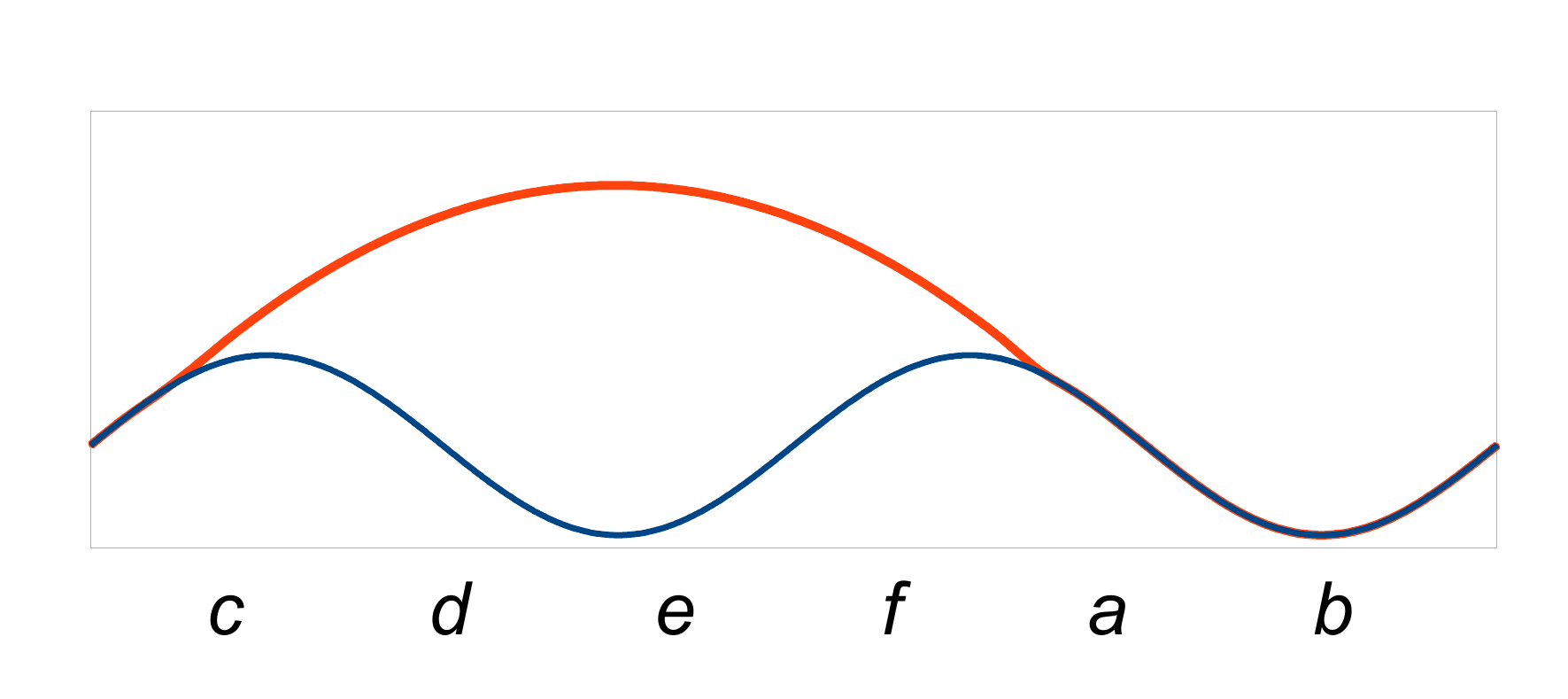}
	\caption{\em The height of the droplet (red) and the surface  
	(blue) as it oscillates vertically with time.
	Surface waves are neglected.
	The labels cdefab refer to the photograph in figure \ref{fig:couder2006prlbounce}.}
	\label{fig:Bouncing-feedback}
\end{figure}

To first order the height of the surface $h$ above the ambient level 
obeys the wave equation
\begin{equation}
	\frac1{c^2} \frac{\partial^2 h}{\partial t^2}~
	-~ \frac{\partial^2 h}{\partial x^2}~
	-~ \frac{\partial^2 h}{\partial y^2}~
	= ~0
\label{eq:wave}
\end{equation}
where $c$ is the wave speed. Figures \ref{fig:couder2006prlbounce} indicates 
that the waves are predominantly standing waves rather than propagating ones
(note the reduced amplitude in photographs b and e).
The relevant circularly symmetric solution is
\begin{equation}
	h~~=~~ -h_o ~\cos(\omega_o t) ~J_o(\omega_o \: r/c)
	\label{eq:bessel-standing-wave}
\end{equation}
where $h_o$ is the maximum height and $J_o$ is a Bessel function of the first kind. 

\subsection{Parametric reinforcement}
\label{sec:parametric-driving}

\begin{figure}[htb]
	\centering
		\includegraphics[width=0.5\textwidth]{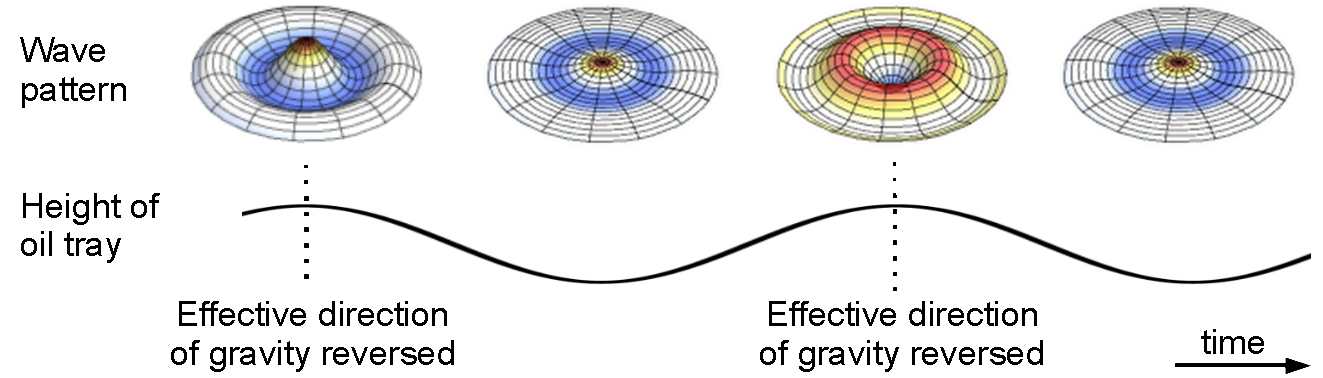}
	\caption{\em Parametric reinforcement of the standing waves in \eqref{eq:bessel-standing-wave}.}
	\label{fig:parametric-driving}
\end{figure}

The standing waves in \eqref{eq:bessel-standing-wave} are reinforced parametrically.
See  Figure \ref{fig:parametric-driving}.
When the oil tray is at its greatest height it is
accelerating downwards, typically at 3.3 -- 4.3 g. 
This reverses the effective force of gravity, 
lifting the wave crests and enlarging the waves.

\begin{figure}[htb]
	\centering
		\includegraphics[width=0.5\textwidth]{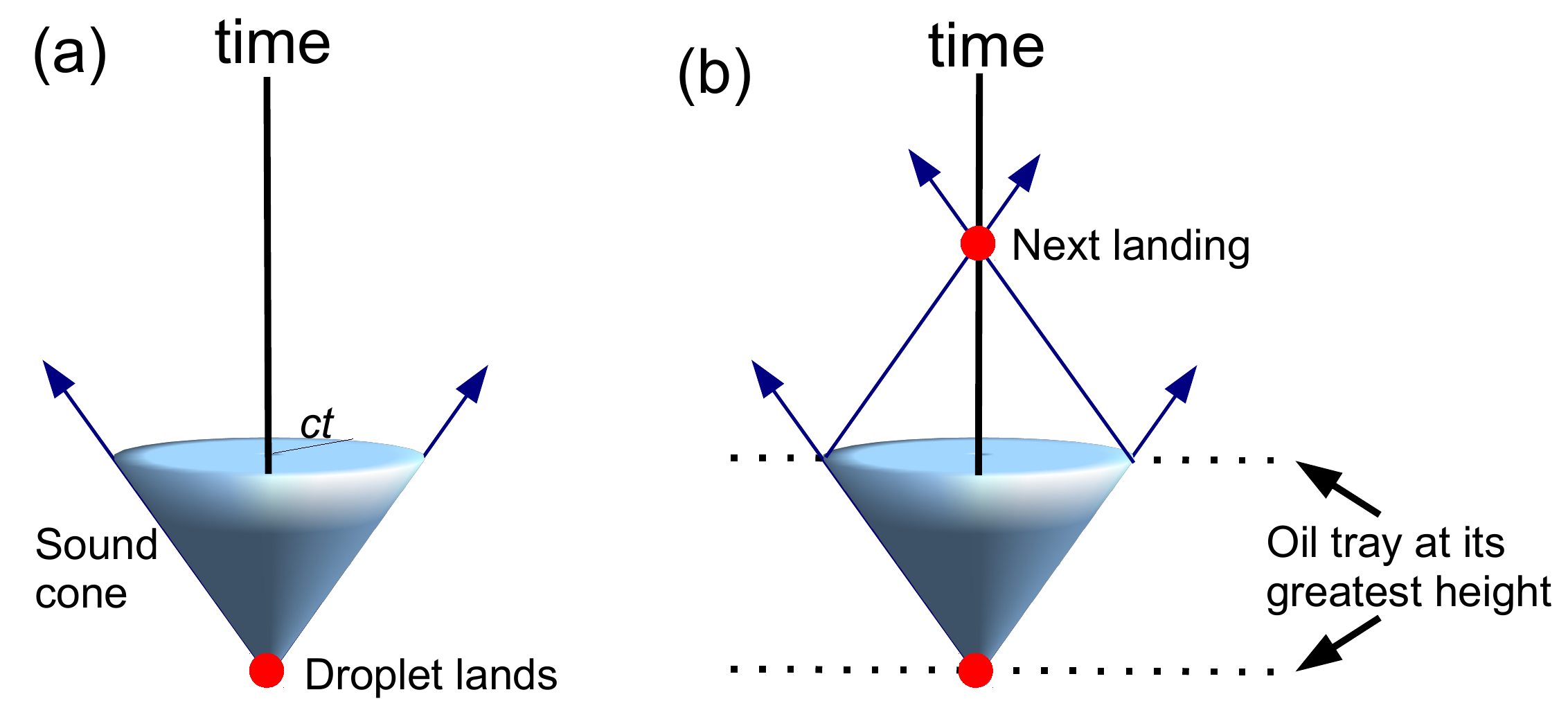}
	\caption{\em Bragg reflection. (a)
When a droplet lands it creates a trough in
the surface which propagates outwards (marked `sound cone'). (b) When
the oil tray is at its greatest height it accelerates downwards. This reverses the
effective direction of gravity, reinforcing the wave and forming an inward-directed 
trough that reaches the centre when the droplet next lands.}
	\label{fig:bragg-mirror}
\end{figure}

This can also be understood in terms of 
propagating waves as shown in figure \ref{fig:bragg-mirror}.
Like a pebble thrown into a pond, a droplet creates outgoing propagating waves when it lands.
They are reinforced and reflected back 
when the oil tray is at its greatest height. The combination of outgoing
and incoming waves forms the standing waves.

When the parametric driving is large enough, 
each bounce of the droplet influences the next through this mechanism. 
The system behaves as if it had
a memory of previous bounces; this is called the `high memory' regime.

\subsection{Wave speed}

The speed $c$ in \eqref{eq:wave} depends on frequency. We do not pursue this 
complication since the frequency was not varied in the experiments. 

Consider an isolated propagating wave.
If it oscillates in-phase with the forcing 
acceleration (similar to figure \ref{fig:parametric-driving}) at one position, 
it will have the opposite phase a quarter of a wavelength away.
Thus, any effect on the wave speed will approximately cancel and the
phase velocity is largely unaffected.

But the standing wave \eqref{eq:bessel-standing-wave} near the droplet is always
in-phase with the forcing acceleration.
The restoring force on the wave is proportional to 
$h \{ g - a_m \cos(2 \omega_o t) \}$ where $a_m$ is the maximum vertical acceleration, so the 
net change of momentum on a half-cycle is proportional to
\[
	\int_{-\frac{\pi}{2}}^{\frac{\pi}{2}} 
	    \cos(\omega_o t) \{g - a_m \cos(2 \omega_o t)\} dt
	\propto g - \frac{a_m}{3}
\]

Since $a_m > 3 g$ in all the relevant experiments,
the parametric driving outweighs the net 
restoring force of gravity and reverses it, leaving only surface tension to
restore the waves. 
This substantially reduces the speed $c$. 
Surface tension is less effective in restoring larger waves, which are slowed the most.  
These standing waves are the ones that interact with the droplet and determine its motion,
so we will need a reduced value of $c$ in the equations that follow.

If the forcing acceleration is increased further, 
eventually it overcomes the restoring force of surface tension and  
$c$ approaches zero, resulting in a Faraday instability~\cite{benjamin1954stability}.

\section{Lorentz symmetry}
\label{sec:walker}

\begin{figure}[htb]
	\centering
		\includegraphics[width=.45\textwidth]{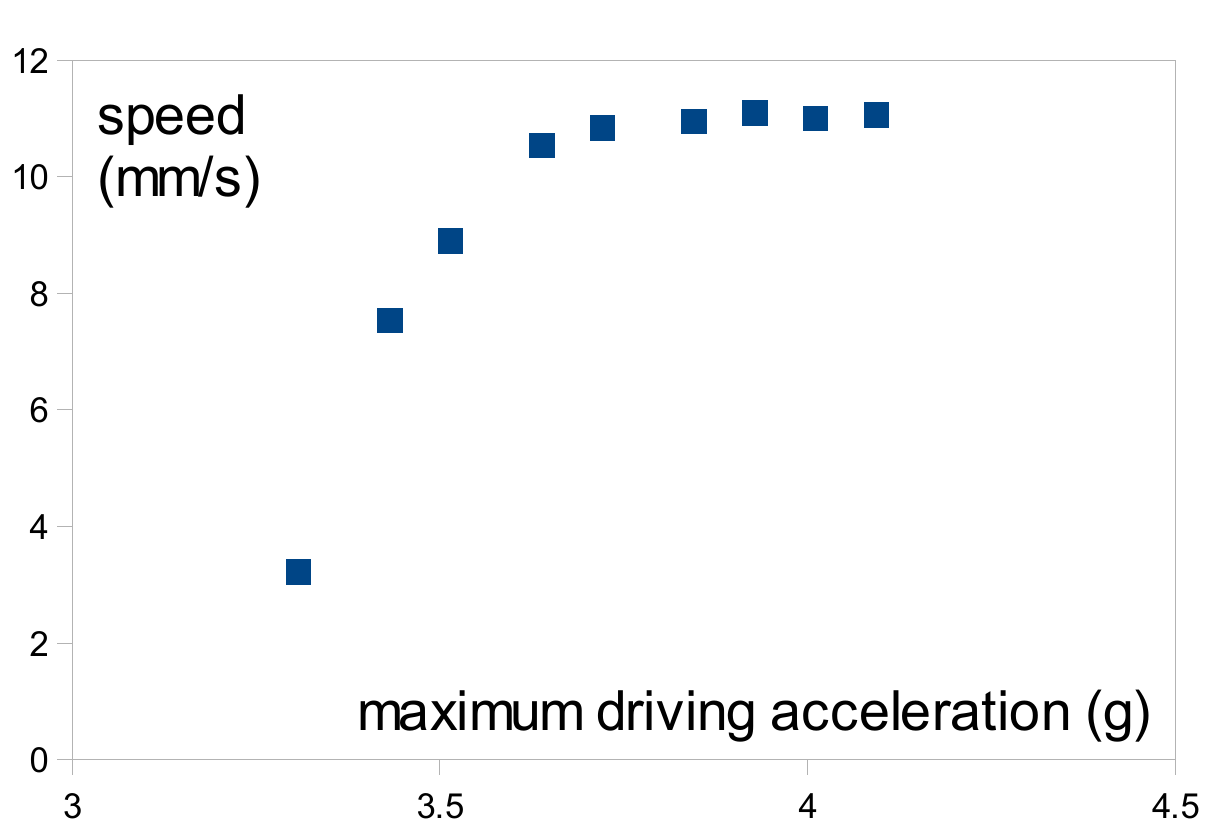}
                \caption{\em The speed of a walker depends on the
                  maximum vertical driving acceleration, graphed as a
                  multiple of the acceleration $g$ due to gravity.
                  (Data courtesy Antonin Eddi, published
                  in~\cite{eddi2011information})}
	\label{fig:eddi2011velocity}
\end{figure}

The droplet in figure \ref{fig:couder2006prlbounce} moves to the right as it bounces. Its 
speed depends on the vertical driving acceleration 
as shown in Figure \ref{fig:eddi2011velocity}. 

\begin{figure}[htb]
	\centering
		\includegraphics[width=0.45\textwidth]{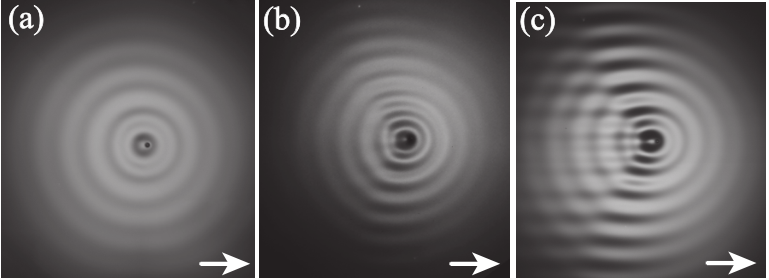}
	\caption{\em A droplet moving to the right. At low speed (left) it has been displaced from the centre; at higher speeds the wave field becomes more complex. (Courtesy Antonin Eddi, Eric Sultan, Julien Moukhtar, Emmanual Fort, Maurice Rossi and Yves Couder~\cite{eddi2011information})}
	\label{fig:low-speed-droplet}
\end{figure}

As we can see in figure \ref{fig:low-speed-droplet}, at low speed the droplet 
merely appears to be displaced slightly to the right, but as the parametric 
driving and the droplet speed increase, the wave field supporting the droplet 
becomes more complex. See Oza, Rosales and Bush for a detailed treatment~\cite{oza2013trajectory}.

\subsection{Simplified model}

We now derive a simplified model of this motion using the symmetries of the
wave equation, and test its predictions against the experimental data. 

If $h(x, y, t)$ obeys the wave equation \eqref{eq:wave}, then so does
$h(x', y', t')$ where

\begin{align}
	x' ~~&=~~ \gamma ~(x - v t) \nonumber\\
	y' ~~&=~~ y \nonumber \\
	t' ~~&=~~ \gamma \left( t - \frac{v x}{c^2} \right) \nonumber \\
	\gamma ~~&=~~ \frac{1}{\sqrt{1 - \frac{v^2}{c^2}}}
	\label{eq:lorentz}
\end{align}

\begin{figure}[htb]
	\centering
		\includegraphics[width=0.45\textwidth]{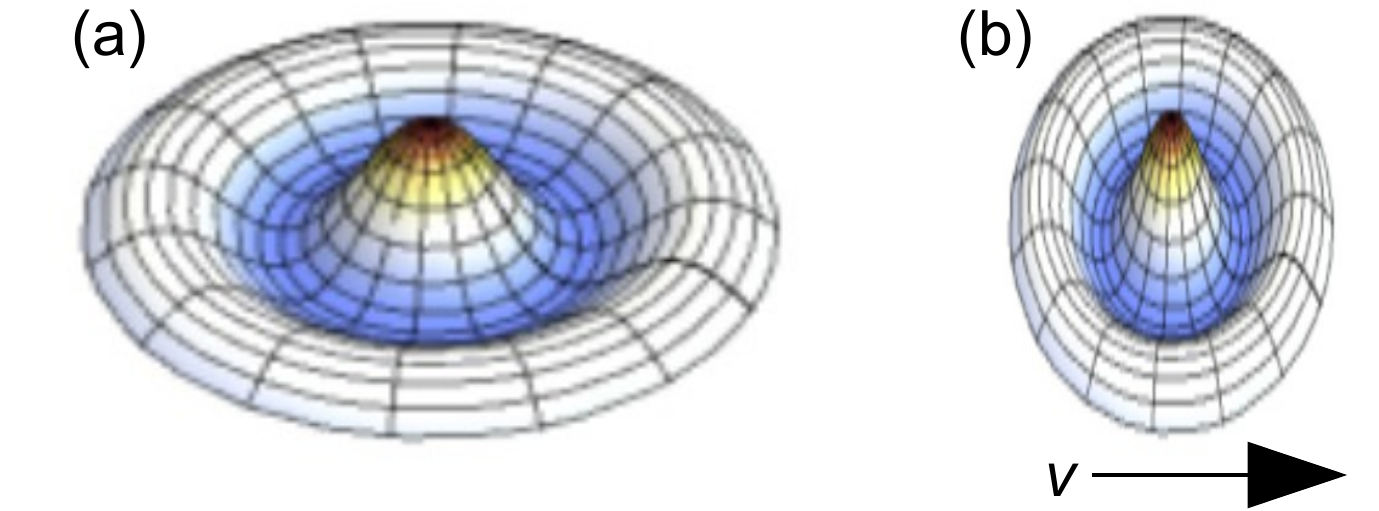}
	\caption{\em (a) The waves near a stationary droplet 
	$\cos(\omega_o t) J_o(\omega_o r/c)$ (b) The Lorentz-boosted wave field
	$\cos(\omega_o t') J_o(\omega_o r'/c)$ moves at speed $v$.}
	\label{fig:j0-lorentz}
\end{figure}

This is the Lorentz transformation familiar from electromagnetism; it also 
applies to solutions of the wave equation in fluids, and is used in aerodynamics
and acoustics (e.g.~\cite{dubois2011}). Applying it to 
\eqref{eq:bessel-standing-wave} for the waves near a stationary droplet 
gives $\cos(\omega_o t') J_o(\omega_o r'/c)$ where $r'^2 = x'^2 + y'^2$, which is 
illustrated in Figure \ref{fig:j0-lorentz}.

In the moving solution, all lengths in the direction of travel
have contracted by the factor $1/\gamma$ (substitute $t = 0$ into
\eqref{eq:lorentz}), while all time periods have dilated by the factor $\gamma$
(substitute $x = v t$). 

The driving frequency is fixed, so the wave field of the droplet must
adapt to compensate. If $h(x', y', t')$ obeys the wave equation 
then so does $h(\alpha x', \alpha y', \alpha t')$ where $\alpha$ is a scale factor. 
Choosing $\alpha = \gamma$ gives
\begin{align*}
	x'' ~~&=~~ \gamma^2 (x - v t) \\
	y'' ~~&=~~ \gamma ~y  \\
	t'' ~~&=~~ \gamma^2 \left( t - \frac{v x}{c^2} \right)
\end{align*}
Applying this coordinate transformation to 
\eqref{eq:bessel-standing-wave} gives
\begin{equation}
	h=-h_o \: \cos \left(\omega_o t -  \frac{\gamma^2 \omega_o  v }{c^2} 
	       \Delta x \right) J_o \left(\frac{\omega_o}{c} r'' \right)
	\label{eq:moving-solution}
\end{equation}
where $\Delta x = x - v t$.
This may be a reasonable approximation to the waves near a walker
because it obeys the wave equation, it moves at speed $v$, 
and it bounces at constant frequency.

As the vertical acceleration increases, the droplet is thrown higher and 
lands later in the cycle (see figure \ref{fig:Bouncing-feedback}). Suppose
it lands at $t = n \tau + T$ where $\omega_o \tau = 2 \pi$.
From \eqref{eq:moving-solution} with $(\Delta x, y, t) =(0, 0, T)$,
\begin{align}
	\frac{\partial h}{\partial x} &= 
	        - h_o ~ \frac{\gamma^2 \: \omega_o}{c} ~ 
			   \left(
			       \frac{v}{c}\: \sin(\omega_o T)
			   \right)
\nonumber
\\  \frac{\partial^2 h}{\partial x^2} &= 
              h_o ~ \frac{\gamma^4 \: \omega_o^2}{c^2} \:
			   \left(
					\frac{v^2}{c^2} \cos(\omega_o T) + \frac12
				\right)
\label{eq:moving-droplet-derivatives}
\end{align}
%

The sloping surface displaces the droplet 
from the centre, as we can see in figure \ref{fig:low-speed-droplet}.
It will settle near $\partial h/\partial x = 0$, where, from 
\eqref{eq:moving-droplet-derivatives},
\[
	\gamma^2 \omega_o \left(\frac{v^2}{c^2} + \frac12 \right) \: \Delta x ~~=~~ v T
\]
where we have approximated $\sin(\omega_o T) =  \omega_o T$ and $\cos(\omega_o T)=1$.
Subsequent waves will always be generated at this displaced position, 
with the net result that the wave pattern moves at speed $v \propto \Delta x$ to
first order, and
\begin{equation}
	\gamma^2 \left( \frac{v^2}{c^2} + \frac12 \right) ~~\propto ~~T
	\label{eq:velocity-prediction}
\end{equation}

\subsection{Comparison with experiment}

\begin{figure}[ht]
	\centering
		\includegraphics[width=0.5\textwidth]{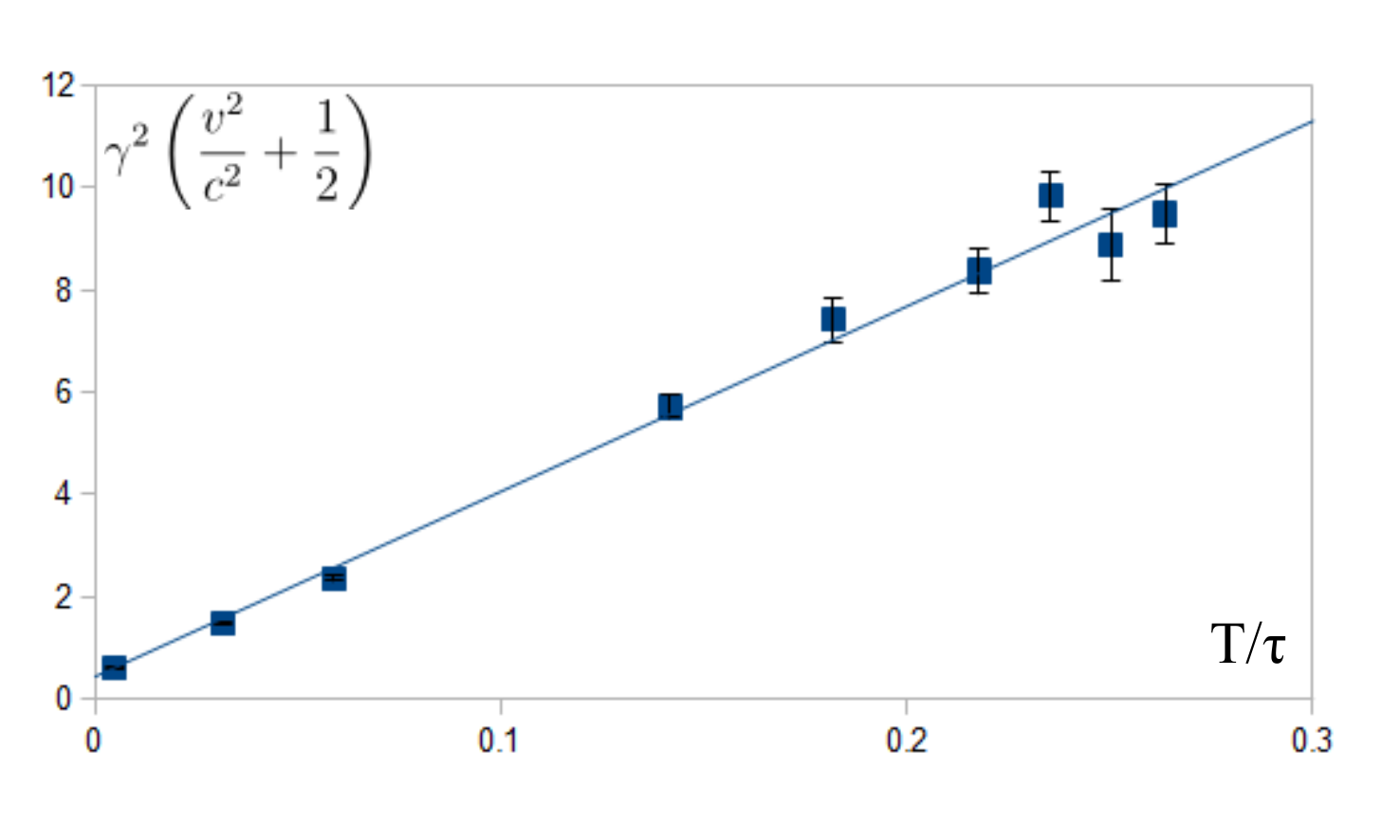}
	\caption{\em The experimental data in figure \ref{fig:eddi2011velocity}
	plotted on new axes. 
	$T$ is the landing time.
	Compare equation \ref{eq:velocity-prediction}.}
	\label{fig:eddi2011tau-gamma}
\end{figure}

The test for such a simplified model can only come from the experimental data,
and we see in Figure \ref{fig:eddi2011tau-gamma} that the linear relationship
we derive between $\gamma^2(v^2/c^2 + \tfrac12)$ and $T/\tau$ is remarkably accurate out 
to an acoustic Lorentz factor of $\gamma = 2.6$.
The landing time $T$ was obtained from the intersection between 
the path of the droplet, for which $\ddot{h} = - g$, and the vertical oscillation
of the oil tray (see figure \ref{fig:Bouncing-feedback} where the maximum
acceleration was 3.5g).
The characteristic speed $c$ of the standing waves near the droplet was
taken to be $11.95$ mm/s, which is 8\% larger than the maximum speed measured. 

\begin{figure}[htb]
	\centering
		\includegraphics[width=0.35\textwidth]{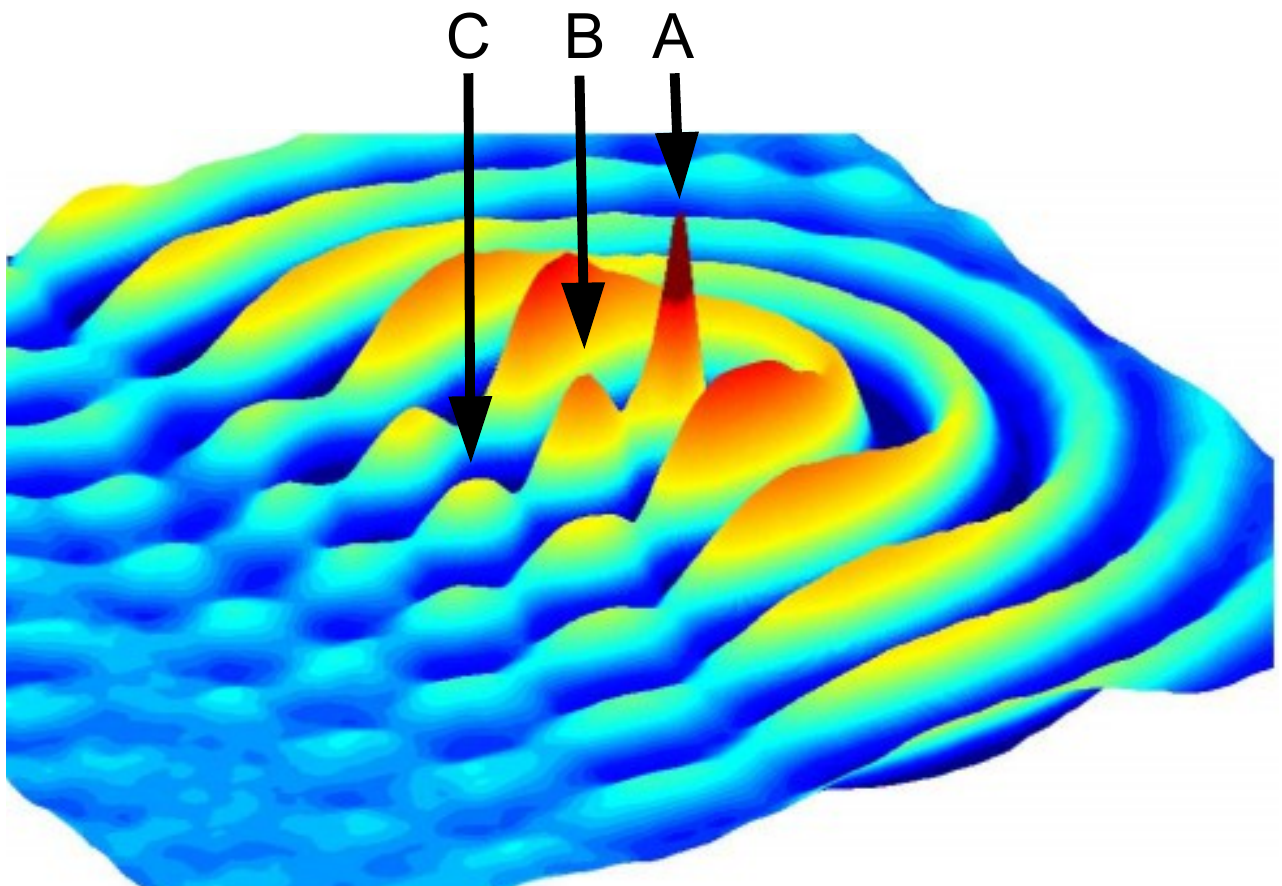}
	\caption{\em The wave field near a walker at large driving amplitude.   
	Courtesy Antonin Eddi, Eric Sultan, Julien Moukhtar, Emmanual Fort, Maurice Rossi and Yves Couder~\cite{eddi2011information}}
	\label{fig:walker}
\end{figure}

There is further information in detailed velocimetry studies, reported by Eddi,
Sultan, Moukhtar, Fort and Couder in~\cite{eddi2011information}. Figure 
\ref{fig:walker} is the wave field measured near a walker at high parametric 
driving. Successive bounces of the droplet can be seen in the peaks marked $A, 
B$ and $C$. We can obtain an approximation to the wave field by treating these 
three peaks 
as the centres of three wave fields given by \eqref{eq:moving-solution}. 
This is approximate because it neglects viscosity and nonlinearities such as those 
introduced by the parametric driving.

The three waves reinforce nearly perpendicular to the direction of motion, 
as can be seen from the taller waves there. They interfere destructively at an angle behind
them, producing the wake-like lines with nearly zero amplitude. 

\begin{figure}[htb]
	\centering
		\includegraphics[width=0.35\textwidth]{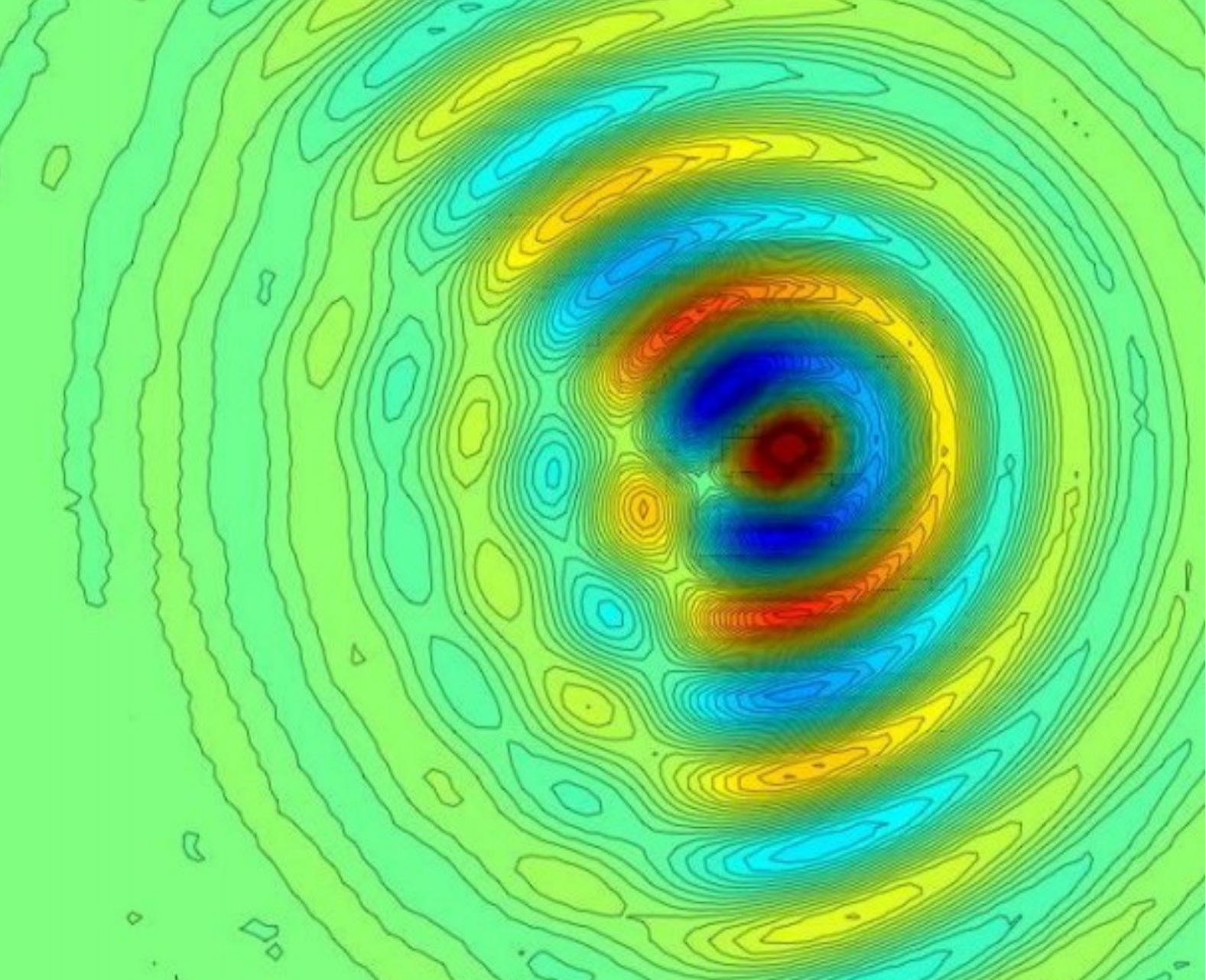}
	\caption{\em Contours of the waves in figure \ref{fig:walker}.}
	\label{fig:walker-2d}
\end{figure}

Taller waves have a reduced wave speed, due to the parametric driving,
which will compress the wave pattern perpendicular to the direction of motion.
At the same time, the wave pattern is elongated parallel to the direction of travel
because the source ($A, B, C$) is elongated.
However, there is a counteracting effect. 
The Lorentz contraction in \eqref{eq:moving-solution} compresses the wave
pattern in the direction of motion. As we can see in figure \ref{fig:walker-2d},
the resulting waves are roughly circular.
See~\cite{eddi2011information,molacek2013dropswalking} for further studies of the wave field.

We conclude that bouncing droplets can be considered to be Lorentz covariant, to
a good approximation, up to a Lorentz factor of 2.6. 

We now turn to experiments conducted at constant forcing frequency and amplitude, 
where the velocity of the droplet is constant and the perturbations 
inflicted by the parametric driving on the Lorentz symmetry can largely be treated as constant and 
neglected.

\section{Force between droplets}
\label{ch:interactions}

\begin{figure}[htb]
	\centering
		\includegraphics[width=0.35\textwidth]{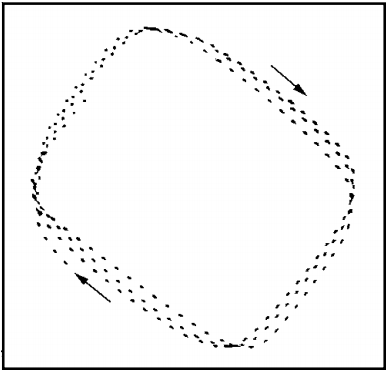}
	\caption{\em Stroboscopic photograph of a droplet's path (dots) deflected near the walls of the container (solid).  (Courtesy Suzie Proti\`ere, Arezki Badaoud and Yves Couder)~\cite{protiere2006particle}}
	\label{fig:protiere2006deflection}
\end{figure}

When a walker approaches the edge of the
container, it does not actually touch the edge but is deflected away.  The
stroboscopic photograph in figure \ref{fig:protiere2006deflection} shows a
droplet travelling three times round a rectangular dish. This experiment gives
deep insight into one type of interaction between droplets.

\subsection{Velocity normal to the boundary}

\begin{figure}[htb]
	\centering
		\includegraphics[width=0.45\textwidth]{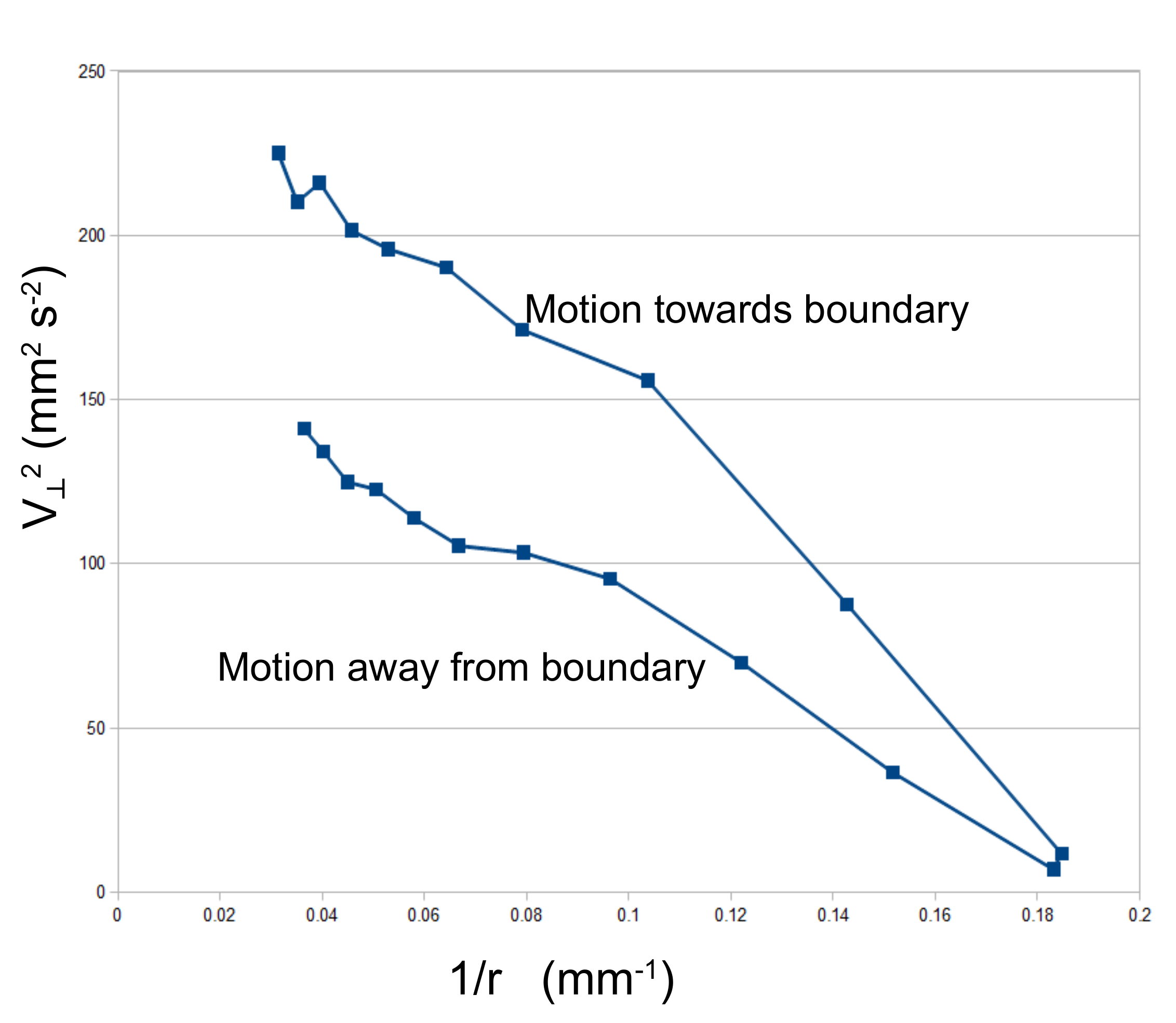}
	\caption{\em The square of the velocity normal to the boundary, $V_\perp^2$, as a function of inverse distance from the boundary. The data are extracted from the stroboscopic images near the bottom of figure \ref{fig:protiere2006deflection}.}
	\label{fig:reflection-velocity}
\end{figure}

The velocity normal to the boundary, $V_\perp$, can be measured from the
photograph in figure \ref{fig:protiere2006deflection}, where an equal time
passes between each stroboscopic image.  Figure \ref{fig:reflection-velocity}
plots $V_\perp^2$ as a function of the inverse distance $\frac1r$ from the
boundary.

For each branch in figure \ref{fig:reflection-velocity}, the data near the
boundary (towards the right of the graph) fall very nearly on a straight line,
before deviating at greater distances. 
This straight line can be written
\begin{equation}
	V_\perp^2 ~~=~~ V_o^2 ~-~  \frac{B}{r}
\label{eq:droplet-repulsion}
\end{equation}
where the slope of the graph is $-B$ and it depends on the branch.  
Extrapolating to $1/r = 0$ on the upper branch gives $V_o \approx ~$18mm/s, which
is the same as the speed of the droplet to the accuracy of measurement.  The
lower branch has $V_o \approx 14 $mm/s.  

\subsection{Inverse square force}



\begin{figure}[htb]
	\centering
		\includegraphics[width=0.45\textwidth]{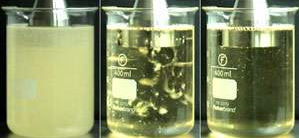}
	\caption{\em Degassing oil by applying ultrasonic vibration. The process takes about 5 seconds. (courtesy Hielscher Ultrasonics GmbH)}
	\label{fig:degassing-oil}
\end{figure}

\begin{figure}[htb]
	\centering
		\includegraphics[width=0.35\textwidth]{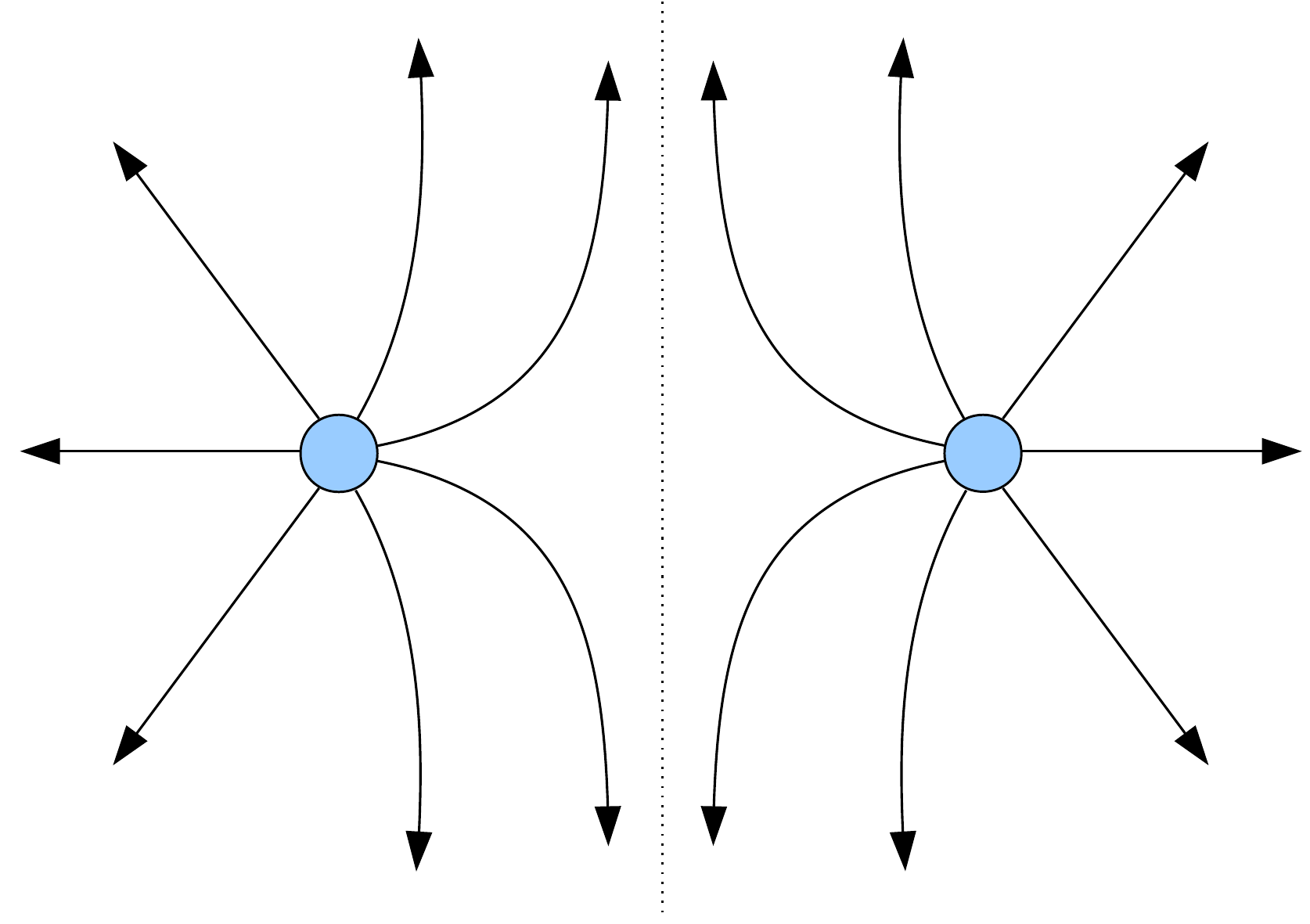}
	\caption{\em Schematic of the flow from two sources.}
	\label{fig:flow-two-sources}
\end{figure}

These experimental results show there is an inverse square force of repulsion
near the boundary. 
The mechanism responsible for it is in fact well
known~\cite{faber1995fluid}. It is used for removing unwanted bubbles of gas from oils and other
liquids using ultrasonic vibration, as in figure \ref{fig:degassing-oil}.
Ultrasonic pressure waves cause nearby bubbles to expand and contract
in phase with one another, inducing
oscillatory radial flows in the liquid.
Near the mirror plane equidistant from two bubbles, the
flows reinforce as illustrated in figure \ref{fig:flow-two-sources}.
The increased velocity results in a reduced Bernoulli
pressure, giving rise to a force of attraction between the bubbles which merge and rise
to the surface.

The exact same mechanism causes a bouncing droplet to avoid the boundary of the
dish, which has the same effect as an imaginary image droplet on the other
side, at the same distance, and bouncing antiphase. Each droplet drives radial
flows in the liquid, just like those near the bubbles in the degasser -- except
that the bouncing droplets are antiphase, so the force is one of repulsion rather than attraction.

\subsection{Size of the bubble force}
\label{magnitude-of-force}

\begin{figure}[htb]
	\centering
		\includegraphics[width=0.35\textwidth]{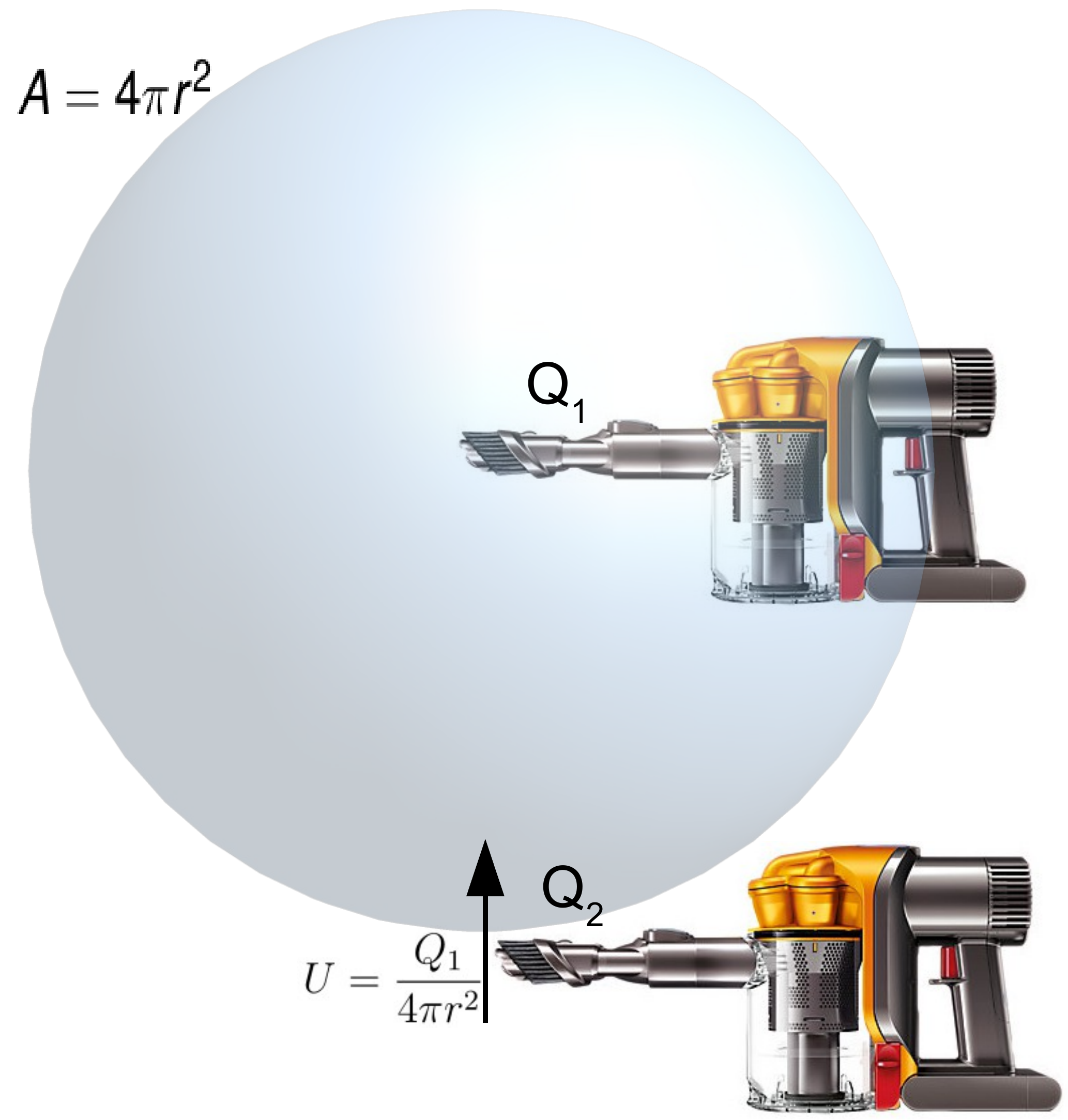}
	\caption{\em The flow near two vacuum cleaner nozzles, neglecting exhaust.}
	\label{fig:dyson3}
\end{figure}

In Figure \ref{fig:dyson3}, a vacuum cleaner nozzle ingests volume $Q_1$ of air per 
unit time. If the flow is spherically symmetric then the air
speed at radius $r$ will be $U = -Q_1 /(4 \pi r^2)$.
A second nozzle at this radius, with volume $Q_2$ per unit time, will ingest
momentum along with the air particles 
\begin{equation}
\frac{dp}{dt}~ =~ \rho_o U Q_2 ~=~ -\rho_o ~\frac{Q_1 Q_2} {4 \pi r^2}
\label{eq:bubble-force}
\end{equation}
where $\rho_o$ is the density. 
This is an inverse square force of attraction
between the nozzles.

The direction of flow, and hence of the force, will be reversed if one of them
is set to blow; then the force will be reversed for both hoses, by conservation
of momentum (we assume the flow remains spherically symmetric, which might be
achieved using a baffle on the end of the hose). More generally, oscillatory
motion results in an attractive force if it is in-phase, and a repulsion if it
is antiphase. The magnitude of the force is given by \eqref{eq:bubble-force}
averaged over a cycle.

\subsection{Size of the droplet force}

In the droplet experiments, the force in \eqref{eq:bubble-force} 
must be doubled due to the hemispherical geometry but halved to average over a cycle,
leaving its magnitude unchanged.

Suppose a droplet of volume $V$ bounces at frequency $f$.
It will induce a flow $f V$ directly, which will be enhanced by 
secondary flows due to entrained fluid and the reinforced waves, 
giving $Q = \beta f V$ where $\beta$
is a factor into which we will also incorporate the effects of 
higher harmonics.
Substituting into \eqref{eq:bubble-force} and remembering to invert the sign, 
the acceleration of a droplet is
\[
	a~=~-\frac{F}{\rho_o V}~=~\frac{V \beta^2 f^2}{4 \pi r^2}
\]
Using $f=25$Hz and a droplet radius of 0.35mm (which was reported for a different run), the acceleration measured in the upper branch of figure \ref{fig:reflection-velocity} gives 
$\beta \approx 5$.

\subsection{Bubble force in conventional form}

An inverse square force can always be written in the form
\begin{equation}
	F ~~=~~ \alpha ~\frac{\bbar c}{r^2}
\label{eq:def-inverse-square-force}
\end{equation}
where $\alpha$ is a dimensionless constant and $\bbar$ is a constant with the
dimensions of energy $\times$ time.

Suppose the radius of a bubble is given by
\[
	r_b~~=~~r_o(1 + A \sin \omega t)
\]
We will simplify the calculation by assuming that $A$ is small. The flow speed
at the surface is
\[
	v_s~~=~~\frac{dr_b}{dt}~~=~~A ~r_o \omega \cos (\omega t)
\]
Multiplying by the area, the flow is $Q = 4 \pi r_o^2 v_s$. Substituting into \eqref{eq:bubble-force} gives
\begin{align*}
	F~~&=~~ 4 \pi \rho_o r_o^3 . r_o^3 A^2 \omega^2 \cos^2(\omega t)  \frac1{r^2}
\\	 ~~&= ~~ 3m_d . r_o^3 A^2 \omega^2 \frac1{2 r^2}
\end{align*}
where  $m_d$ is the displaced mass of the bubble
and we have replaced $\cos^2(\omega t)$ by its average value, $\frac12$. 

This can be rearranged into the conventional form 
\eqref{eq:def-inverse-square-force}
using the fact that the inertial mass of the bubble, 
due to the motion of the displaced fluid around it, is 
approximately $m = \frac12 m_d$~\cite{faber1995fluid}. Thus
\begin{align}
	\alpha ~~&=~~ 3 A^2~ \left( \frac{r_o \omega}{c} \right)^3 \nonumber
\\	\bbar~~&=~~\frac{m c^2}{\omega} 
\label{eq:define-bbar}
\end{align}

The dimensionless constant $\alpha$ depends on whether the bubbles are resonant
or not.  Consider the resonant case.  Neglecting geometric factors (which are
of order 1), the bubble radius will vary from a small value to $2 r_o$, giving
$A \approx 1$.  The maximum surface speed $r_o \omega$ will also increase, but
it cannot much exceed the speed of sound in the fluid since the pressure
would reduce to zero due to the Bernoulli effect.  Therefore, if the bubbles
are resonant then both $A$ and the ratio $r_o
\omega_o/c$ will be of order 1, and so $\alpha$ is also of order 1.

\subsection{Constant of the motion}
\label{sec:bbar-constant-of-motion}

If an unperturbed acoustic Lorentz transformation \eqref{eq:lorentz} 
could be realized experimentally,
$\bbar$ would be a constant of the motion.  
The effective mass $m$ of a wave is proportional to its volume, 
so the Lorentz contraction multiplies it by $1/\gamma$,
whilst the angular frequency is multiplied by the same factor, 
so $\bbar \propto m/\omega$ is independent of velocity.
More formally, its dimensions, energy $\times$ time, are Lorentz invariant.

The droplet's speed was not varied during the experiments.
One way to achieve this might be 
to adjust the forcing amplitude and frequency (correcting
for the perturbation to the wave speed and height). 
Alternatively a droplet of ferrofluid might be de-weighted
magnetically so it lands later in the cycle and travels faster.
The forcing frequency might be adjusted 
to avoid the additional factor $(1 - v^2/c^2)$ from the scale enlargement 
in \eqref{eq:moving-solution}.
The mass of the droplet itself must be treated separately from that of the waves.

\subsection{Comparison to the force between electrons}

The electrostatic force between electrons is
\begin{align*}
	F~~&=~~\alpha ~\frac{\hbar c}{r^2}
\\	\alpha ~~&\approx ~~\frac1{137.036}
\\	\hbar~~&=~~\frac{m c^2}{\omega}
\end{align*}
where $m$ is the mass of the electron, $\omega$ its angular frequency, $c$ is
the speed of light, and $\hbar$ = $h/2\pi$ where $h$ is Planck's constant.

Notice the analogy to \eqref{eq:def-inverse-square-force} and \eqref{eq:define-bbar},
where $\bbar$ is defined in the same way as $\hbar$.

From the value of $\alpha$, it seems that the electrostatic force is about two orders of magnitude weaker
than the mechanical force between resonant bubbles. This suggests one
limitation of the bouncing-droplet experiment as a model of quantum mechanics,
namely that spherically-symmetric resonant solutions are not a good model for
the electron.  We will explore higher-order solutions that are not spherically
symmetric below.

\subsection{Maxwell's equations}
\label{maxwell-equations}

When the droplets or bubbles are stationary, we have seen there is an inverse square 
force between them (when averaged over a cycle). This obeys the same equations 
as the electrostatic field near a charged particle, since both are inverse square. 

These equations can be extended to the case of moving droplets
by noting that the solutions are acoustically Lorentz covariant to a reasonable approximation.
So we need equations that are Lorentz covariant and that reduce to the equations of
electrostatics when stationary.

These conditions are met by Maxwell's equations with an acoustic value of $c$. 
In fact, they are unique in that Maxwell's equations (more
strictly, equations that are equivalent to them when they are averaged over a
cycle) are the only ones that satisfy them. Suppose the contrary, that there
existed a different set of Lorentz covariant equations that produce the same
electric field with the same boundary conditions. The only difference between
the two solutions can be in the magnetic field. But a Lorentz transformation
turns a pure magnetic field into one with an electrical component, and so the
electrical fields differ in the new reference frame. This is a contradiction.
Thus the interaction between the droplets obeys Maxwell's equations when it is
averaged over a cycle. 

This model predicts an interaction obeying the equations of magnetism, 
which is observed experimentally as follows.

\subsection{Magnetic interaction}
\label{sec:magnetic-interaction}

We now turn to the lower branch of figure \ref{fig:reflection-velocity}.

A walker's speed is fixed by the driving amplitude as discussed above. 
As the walker's velocity normal
to the boundary slows down and reverses in the experiment, it must accelerate parallel to the
wall to maintain constant speed. The velocity boost is observed in the experiment.
The researchers estimated the angle of incidence (relative to the normal to the
boundary) at about $38^\circ$, and of reflection at about $53^\circ$. 

In addition to the force of repulsion between the droplets, 
which obeys the same equations as those of electrostatics,
there is a force of attraction when they are moving at a common velocity $v$
parallel to the boundary. This obeys the equations of magnetism. 
It is like the magnetic force of attraction between two electrons 
moving at a common velocity parallel to one another.

The magnetic force reduces the total force by a factor $1 - v^2/c^2$,
which accounts for the reduced slope of the lower graph 
in figure \ref{fig:reflection-velocity}.
We will take the approximation that 
the upper branch in figure \ref{fig:reflection-velocity} has no velocity perpendicular to the
direction of travel, but by the time the droplet has reached the lower branch
it has been accelerated to the full perpendicular speed, $v$ = 18 mm s$^{-1}$
by the tangential force. The force is proportional to the slope of the graph,
whose ratio is 14/18. Equating these two gives $v = 0.47 c$, which suggests
the droplets were moving at about half the wave speed.

\subsection{Propagating waves}
\label{sec:propagating-waves}

Maxwell's equations tell us that if a source is accelerated, propagating waves
will be emitted. We are not aware of attempts to test for these waves
experimentally with droplets. This might be possible by accelerating 
droplets of a ferrofluid horizontally using magnetics. The period
of the acceleration should be longer than that of the droplets. 

We predict that propagating waves will be observed that are modulations of
the standing waves surrounding the source (not ordinary
longitudinal or transverse waves). They obey Maxwell's equations
with an acoustic value of $c$.

The force obeying Maxwell's equations is only one of the interactions between 
droplets. To see another we must turn to coherent motion.

\section{Diffraction}
\label{sec:diffraction}

\begin{figure}[htb]
	\centering
		\includegraphics[width=0.30\textwidth]{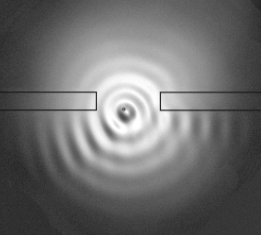}
	\caption{\em A droplet passing through an aperture in a submerged barrier.}
	\label{fig:couder2006one-slit-photo}
\end{figure}

We saw how a droplet is repelled from a barrier.
When the barrier has one or more slits in it, some droplets pass through
as we can see in figure \ref{fig:couder2006one-slit-photo}.

\begin{figure}[htb]
	\centering
		\includegraphics[width=0.3\textwidth]{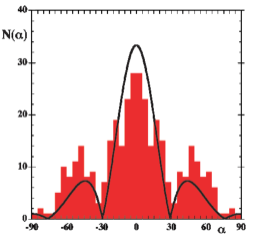}
                \caption{\em Histogram showing the
                  number $N$ of droplets (out of 125) that emerge at angle
                  $\alpha$ to the normal at large distance. The solid line is a 
				  single-slit diffraction pattern. Courtesy Yves
                  Couder and Emmanuel Fort~\cite{couder2006single}.}
	\label{fig:couder2006oneslit}
\end{figure}

When the Paris team measured which direction they went, 
they found classical diffraction patterns as you see for light waves, water
waves or quantum mechanical particles, as in figure \ref{fig:couder2006oneslit}.
Diffraction is only observed in the high memory regime. In the low memory regime
the waves from a droplet have little effect because they propagate away and are 
lost to viscosity.

In these experiments the forcing frequency and amplitude, and hence the
walker velocity $|v|$, were kept constant.
The variation was in $v_x$ and $v_y$.

\subsection{Wavelength}

We saw (equation \ref{eq:bessel-standing-wave}) that the surface height near a
stationary droplet has two component factors which we will now write $\psi$ and
$\chi$ where
\begin{align}
	h~~&=~~\psi ~ \chi \nonumber \\
	\psi ~~&= ~~ \cos(-\omega_o t) \nonumber \\
	\chi ~~&=~~ ~ -h_o ~J_0(k_r r) 
	\label{eq:define-h-psi-chi}
\end{align}


We are interested in how the wave field of a moving droplet varies in the $x$ direction,
neglecting $v_y$. An acoustic Lorentz transformation \eqref{eq:lorentz}
gives the approximate solution
\begin{align*}
	\psi ~~&= ~~ \cos(-\omega_o t')\\
	\chi ~~&=~~ ~ -h_o ~J_0(k_r r')
\end{align*} 
where for simplicity we have neglected the scale enlargement in \eqref{eq:moving-solution},
which is just a constant since $|v|$ is constant.
In this moving solution, the wave field $\chi$ advances with the
droplet at speed $v_x$, and $\psi$ has become a planar wave which
can be written
\begin{equation}
	\psi ~~=~~ \cos(k x - \omega t)
\label{eq:psi-moving-droplet}
\end{equation}

The values of $k$ and $\omega$ can be obtained by defining $S = - \omega_o t'$
and noting, from \eqref{eq:lorentz},

\begin{align}
    k~~&=~~\frac{\partial S}{\partial x}
  	  ~~=~~\frac{\partial S}{\partial t'}\frac{\partial t'}{\partial x} 
	  ~~=~~\frac{\gamma \omega_o}{c^2}~ v_x
	\nonumber \\
	\omega~~&=-\frac{\partial S}{\partial t}
		  ~~=-\frac{\partial S}{\partial t'}\frac{\partial t'}{\partial t} 
		  ~~=~~\gamma \omega_o 
	\label{eq:omega-and-k}
\end{align}

The wavelength of $\psi$ is $\lambda = 2 \pi/k$, or
\begin{equation}
	\lambda ~~=~~  \frac{2 \pi c^2}{\omega v_x} ~~=~~ \frac{b}{p}
\label{eq:droplet-de-broglie-wavelength}
\end{equation}
where $p = m v_x$ is the momentum of the wave and $b = 2 \pi
\bbar$ where $\bbar = m c^2/\omega$. 

If we choose $m$ to be the effective mass of the wave (as discussed 
in section \ref{sec:bbar-constant-of-motion}), then the parameter $\bbar$ that determines 
the wavelength of a droplet is the same as 
the constant in \eqref{eq:define-bbar} that determines its deflection from the boundary
by the inverse square force.

Equation \eqref{eq:droplet-de-broglie-wavelength} is the same as the de 
Broglie wavelength of a quantum mechanical
particle with $b$ instead of Planck's constant $h$.
It can be extended to arbitrary axes.  If the velocity in the direction of interest is
${\bf v} = (v_x, v_y)$ then $\psi$ in \eqref{eq:psi-moving-droplet} becomes
\begin{equation}
	\psi~~=~~\cos({\bf k}.{\bf x} - \omega t)
\label{eq:vector-bessel-wave}
\end{equation}
where
\begin{equation}
	{\bf p}~~=~~\bbar {\bf k}
\label{eq:p-k}
\end{equation}
and ${\bf p} = (p_x, p_y)$ is the momentum.

%
%

\subsection{The diffraction pattern}

We now show that the histogram in figure \ref{fig:couder2006oneslit} agrees
with the diffraction pattern of $\psi$ through the aperture.

The width of the aperture was 14.8 mm and, based on measuring the vertical
distance from the barrier to the first node (measured near the corner to the
right of the aperture), the wavelength of $\psi$ near the aperture was $\lambda
= $~7.3 mm.  When waves of wavelength $\lambda$ diffract through a single
aperture of width $L$, the first minimum of amplitude is at angle $\theta$
where $\lambda = L \sin \theta$.  The above measurements predict this will
occur at $\theta = 30^\circ$.  The minimum in the experimental histogram occurs
between $30^\circ$ and $35^\circ$.

So we observe that the minimum in the histogram occurs where the waves of
$\psi$ interfere destructively.  It is as if the droplet were repelled from
these regions.  This can be understood as follows.  Bigger waves have deeper
wave troughs, so a droplet bouncing in them will be physically lower than one
bouncing elsewhere.  Consequently it will be attracted towards them by the
force of gravity.  Thus droplets move away from the regions of destructive
interference of $\psi$ where the waves are smaller. 

\subsection{Double-slit diffraction}

In another experiment, the droplet was made to diffract through two slits.
\begin{figure}[htb]
	\centering
		\includegraphics[width=0.4\textwidth]{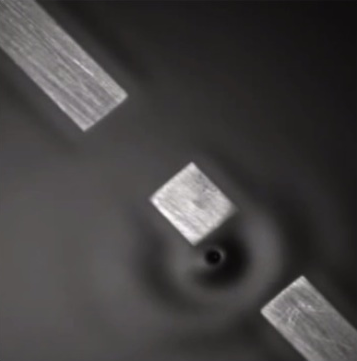}
	\caption{\em Droplet passing through a double slit 
	}
	\label{fig:fort2006twoslit}
\end{figure}

\begin{figure}[htb]
	\centering
		\includegraphics[width=0.4\textwidth]{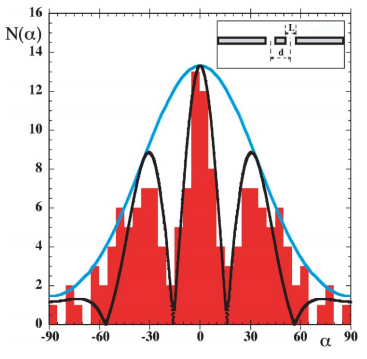}
	\caption{\em Histogram of the deflection angle for 75 droplets that have passed through one of two slits. The solid lines show a possible fit to a double-slit diffraction pattern. Courtesy Yves Couder and Emmanuel Fort~\cite{couder2006single}.}
	\label{fig:couder2006twoslit}
\end{figure}

The histogram of the directions taken by the droplets after they had passed 
through one or other of the slits is shown in figure 
\ref{fig:couder2006twoslit}. 

The distance between the slits was 14.3mm.  Using the above parameters we would
expect the first diffraction minimum to be at approximately $15^\circ$, 
and this is what the researchers observed.

\subsection{Classical approximation}

Defining a quantity with the dimensions of energy
\[
	E~~=~~\bbar \omega
\]
then, from \eqref{eq:omega-and-k},
\begin{align*}
	\omega^2 - c^2 k^2~~&=~~\omega_o^2 \\
	E^2 - p^2 c^2 ~~&=~~m_o^2 c^4
\end{align*}
where we have used $p^2 = \bbar^2 k^2$ from \eqref{eq:p-k} and the definition of 
$\bbar$ in \eqref{eq:define-bbar}.
This will be recognised as the relativistic equation of motion for a
classical particle of rest mass $m_o$ and energy $E$. It has a low-velocity approximation
\[
	E ~~=~~E_o \left( 1 + \frac{p^2 c^2}{E_o^2}  \right)^\frac12
	  ~~\approx~~ E_o + \frac{p^2}{2 m_o}
\]
which is the Newtonian equation of motion.


In order to include the inverse square
force discussed above, it suffices to add a term to the energy, namely
\begin{equation}
	\bbar \omega ~~=~~E~~=~~m c^2~-~V
	\label{eq:define-potential-energy}
\end{equation}
where $V$ is the potential energy associated with the interaction.

Revisiting the experimental constraints, 
in principle it might be possible to adjust the forcing frequency 
in accordance with \eqref{eq:define-potential-energy} during the experiment, but this was not done.
The fixed forcing frequency constrained $|v|$ to be constant.
As we saw in figure \ref{fig:protiere2006deflection}, 
the velocity perpendicular to the direction of interest 
will adjust to compensate.
The experimenters observed an increased tangential speed when the droplets
were near the diffraction slits, but the perturbation to the expected diffraction patterns,
if any, was small.


\subsection{Klein-Gordon equation} 

From \eqref{eq:define-h-psi-chi}, the factor
$\psi$ for a stationary particle obeys the equation
\begin{equation}
	\frac{\partial^2 \psi}{\partial t^2}~~=~~-\omega_o^2 \psi
\label{eq:psi_o}
\end{equation}

In order to extend this to the case of a moving particle,
we need a Lorentz covariant equation
that reduces to \eqref{eq:psi_o} 
in the stationary case, namely
\begin{equation}
	\frac{\partial^2 \psi}{\partial t^2}~-~c^2 \nabla^2 \psi
	~~=~~\omega_o^2 \psi
\label{eq:klein-gordon}
\end{equation}
since the left hand side is Lorentz invariant. 

Equation \eqref{eq:klein-gordon} is the same as the
Klein-Gordon equation of quantum mechanics for a relativistic 
particle. The only difference is
that the characteristic speed in the experiment is
the speed of surface waves in the oil rather than the speed of light.

\subsection{Schr\"odinger equation}

If \eqref{eq:define-h-psi-chi} receives a Lorentz boost with a small velocity 
$v$ in the $x$ direction then we get 
$\psi = \cos(-\omega_o t') = \cos(v x \omega_o/c^2 - \omega_o t)$ where
we have approximated $\gamma = 1$. Writing this in the form
\begin{equation}
	\psi~~=~~R \cos(\theta - \omega_o t)
	\label{eq:define-rs}
\end{equation}
gives $\theta = v x \omega_o/c^2$.
Extending to arbitrary axes gives the velocity of the droplet as 
determined by the local waves
\begin{equation}
	{\bf v}~~=~~\frac{c^2}{\omega_o} \: \nabla \theta
	\label{eq:dbb-velocity}
\end{equation}

The function in \eqref{eq:define-rs} can be analytically continued into the complex 
plane by defining
\begin{equation}
	\psi_s ~~=~~ R\: e^{i \theta}
	\label{eq:define-psi-s}
\end{equation}
so that $\psi = \Re(e^{-i \omega_o t} \psi_s)$ where $\Re$ means the real part.
Now, $\psi$ obeys the Klein-Gordon equation  \eqref{eq:klein-gordon}, 
and we will seek a solution where both the real and imaginary parts of 
$e^{-i \omega_o t} \psi_s$ obey this same equation, which is satisfied when
\begin{equation}
	i \: \frac{\partial \psi_s}{\partial t}
	~~=~~ -\frac{c^2}{2 \omega_o} \: \nabla^2 \psi_s
\label{eq:schrodinger-nohbar}
\end{equation}
where we have neglected the term in $\partial^2 \psi_s/\partial t^2$, which is small
when the velocity is small. 

Substituting \eqref{eq:define-potential-energy} in the form 
$\bbar \omega_o = m_o c^2 - V$ gives
\begin{equation}
	i \:\bbar \: \frac{\partial \psi_s}{\partial t}
	~~=~~\left( -\frac{\bbar^2}{2 m_o} \nabla^2 + V \right) \psi_s
\label{eq:schrodinger}
\end{equation}

This is the same as the Schr\"odinger equation for the wavefunction of 
a quantum mechanical particle.
The only difference is that the constant
of motion in the experiment is $\bbar$ rather than Planck's reduced constant 
$\hbar$ (although they are defined in the same way and they are both 
constants of the motion).

%

\subsection{Probability density}

If the starting position of a droplet is not known precisely,
and it is allowed to evolve over time, then
there will be a range of final positions, which can be calculated probabilistically.
Our treatment will follow the reasoning of David Bohm~\cite{bohm1952suggested}, 
who solved this problem in 1952.

Substituting the definition $\psi_s = R e^{i \theta}$ (equation \ref{eq:define-psi-s})
back into \eqref{eq:schrodinger-nohbar}, and taking
the imaginary part when $\theta=0$ gives
\[
	\frac{\partial R}{\partial t} 
	~~=~~-\frac{c^2}{2 \omega_o} ~ (R \:\nabla^2 \theta - 2 \:\nabla R \:\nabla \theta) 
\]
which can be rearranged into
\begin{equation}
	  \frac{\partial R^2}{\partial t}
	~+~ \nabla (R^2 ~{\bf v})
	~~=~~0
\label{eq:continuity-psi}
\end{equation}
where {\bf v} is the velocity of the droplet in \eqref{eq:dbb-velocity}.

\begin{figure}[htb]
	\centering
		\includegraphics[width=0.50\textwidth]{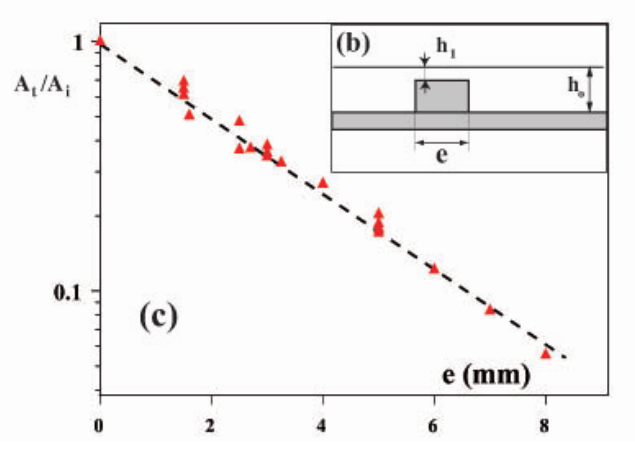}
	\caption{\em Droplets encounter a region of reduced depth, which repels them.
	The $x$ axis is the width of the barrier, and the $y$ axis is the probability 
	the droplet tunnels through the barrier, plotted on
	a logarithmic scale. Courtesy Antonin Eddi~\cite{eddi2009unpredictable}}
	\label{fig:eddi2009tunnelling}
\end{figure}

This equation has a simple interpretation.
When the velocity ${\bf v}$ of a compressible fluid, such as the air, varies with position, its density $\rho$ obeys the continuity equation 
$\frac{\partial \rho}{\partial t} + \nabla (\rho ~{\bf v})=0$,
which is the same as \eqref{eq:continuity-psi} with $R^2$ replaced by $\rho$. Since the
velocity of the droplets is {\bf v}, it follows that
the probability density for the position of the droplet, averaged over
nearby trajectories, must be $R^2 = |\psi_s|^2$ 
(provided the initial value of $|\psi_s|^2$ is appropriately calibrated, or `normalised'). 
This is confirmed by the experimental results in figure \ref{fig:eddi2009tunnelling}.
The graph shows that the probability a droplet crosses the barrier 
(or `tunnels') reduces exponentially with its width.
If you solve Schr\"odinger's equation with a barrier, you get the same
exponential decay of $|\psi^2|$ with the width of the barrier. 
The same probability density $|\psi_s^2|$
is assumed as a postulate in the Copenhagen 
interpretation of quantum mechanics.

The foregoing calculation was performed in 1952, 
more than 50 years before the first droplet experiments.
It led Bohm to hypothesise the existence of a tiny particle which moves at 
the velocity {\bf v} in \eqref{eq:dbb-velocity}, guided by 
waves that obey Schr\"odinger's equation \eqref{eq:schrodinger-nohbar}, whose probability density
is $|\psi_s^2|$. These are exactly the equations for a bouncing droplet at low velocity.
His insight is remarkable. For him, this was a purely abstract exercise; 
he did not have the droplet model to inspire him to derive these relationships from Euler's equation.

Based on these equations, Bohm showed the resulting mechanics to be
indistinguishable from the Copenhagen interpretation of quantum mechanics. He subsequently 
found that Louis de Broglie had suggested 
a similar idea in the 1920s; the model is now called the de Broglie-Bohm
interpretation of quantum mechanics.

\subsection{Same equations, same solutions}

Given that we have the same mathematics up to a constant factor, we expect the
calculations of quantum mechanics to carry over to other droplet
experiments, and can start to understand why bouncing droplets are a pretty good
model of the quantum world.

\begin{figure}[htb]
	\centering
		\includegraphics[width=0.4\textwidth]{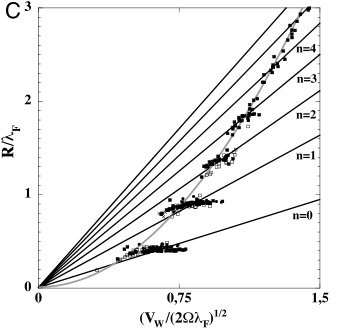}
                \caption{\em A droplet in a rotating bath is attracted towards
                  the centre, and exhibits quantized orbits. Courtesy Emmanual
                  Fort~\cite{fort2010path}}
	\label{fig:fort2010quantisation}
\end{figure}

In figure \ref{fig:fort2010quantisation}, the experiment was conducted in a
rotating bath, where the droplet was attracted towards the centre. 
The droplets exhibited quantized orbits.

Other experimentalists have discovered further evocative results. For example,
Valeriy Sbitnev has recently reported that a droplet and an antidroplet (a
bubble) can be created when two suitably-shaped soliton waves collide on the
surface of the fluid~\cite{sbitnev2013droplets}. The droplet and antidroplet
move apart, just as when a particle and antiparticle are
created in quantum mechanics. However, before we approach the domain of field
theory, we first have to discuss spin.

\section{Rotational motion}
\label{sec:rotational-motion}

\begin{figure}[ht]
	\centering
		\includegraphics[width=0.45 \textwidth]{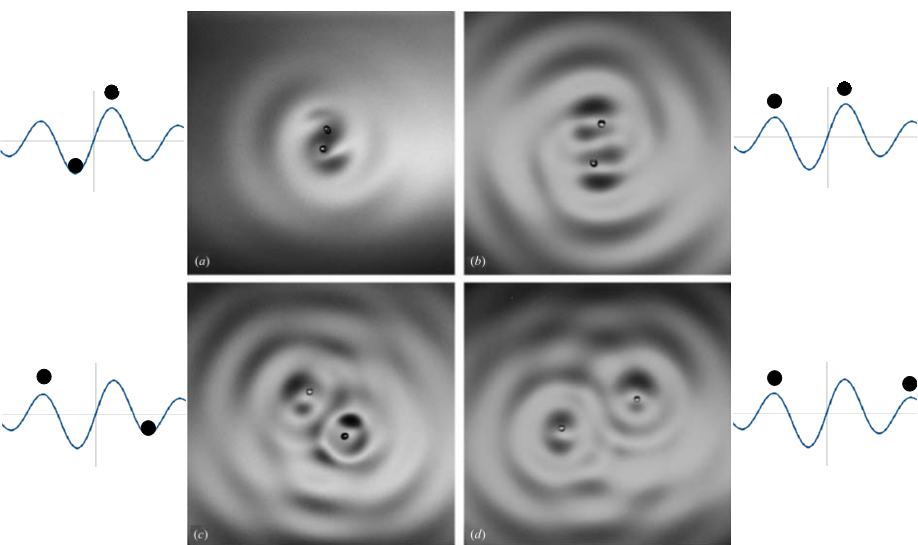}
                \caption{\em The waves with two droplets.  The side drawings
                  show a Bessel function $J_1$, which is the lowest rotating
                  component of the standing waves between the droplets.  The
                  bouncing is antiphase in (a) and (c) and in sympathy in (b)
                  and (d).  (Photograph courtesy Suzie Proti\`ere, Arezki
                  Boudaoud and Yves Couder \cite{protiere2006particle})}
	\label{fig:j1-droplets}
\end{figure}

We have so far only considered waves with circular and spherical symmetry. 
The photographs in figure \ref{fig:j1-droplets} suggest we also have to
think about solutions to the wave equation which depend on angle.

In this experiment, the droplets orbit around one another with a period of
approximately 20 bouncing periods in (a), and longer in (b)-(d).  Their
velocity is approximately that of an ordinary walker driven with the same
vertical acceleration, and the period increases with radius.

\subsection{Harmonic solutions}

The motion in the photographs can be described using the solutions to the
wave equation in circular coordinates $(r, \theta)$, namely
\begin{equation}
	h_m~~=~~h_o ~\cos(\omega_o t - m \theta) ~J_m(k_r r) 
\label{eq:jm-cos}
\end{equation}
where $J_m$ is a cylindrical Bessel function of the first kind, $m$ is an
integer whose sign is significant, and $\omega_o=c ~k_r$.
The waves near the two droplets contain components with various values of $m$,
but the main experimental results can be understood from the lowest order
rotating
components, with $m = \pm 1$. We neglect higher harmonics as well as $J_o$  
in (b) and (d).

Figure \ref{fig:j1-droplets} shows the Bessel function $J_1$,
which gives the wave height of the lowest rotating component
on a line joining the droplets. 
As we have seen, the droplets prefer to land in
the wave troughs, so they are in free flight over the crests, as shown. 
The rotating wave pattern is
\begin{align}
	h~~=&~~\tfrac12 h_0 \: [
				      \cos(\omega_o t + \Omega t - \theta) ~J_1 (k_1 r) \nonumber \\
	               ~&+~ \cos(\omega_o t - \Omega t + \theta) ~J_1(k_2 r)
				]
				\nonumber
\\				    \approx& ~~h_o \: \cos (\omega_o t) ~ 
					           \cos (\Omega ~t - \theta)
							   ~J_1(k_r r)
\label{eq:rotating-droplets}
\end{align}
where $c k_1 = \omega_o + \Omega $ and $c k_2 = \omega_o - \Omega$.  In the
second expression we have used the identity $\cos(A+B) + \cos(A-B) = 2 \cos A
\cos B$, and have approximated $k_1 \approx k_2 \approx k_r$, which is valid at
small $r$ and $\Omega$. This can be regarded as a standing wave that rotates
with the droplets at angular frequency $\Omega$.

The factor $\cos(\Omega t - \theta)$ vanishes on the node line $\theta = \Omega
t \pm \frac12 \pi$.  On either side of this line, its sign reverses; we see in
the photographs in figure \ref{fig:j1-droplets}(b) -- (d) that the crests turn
into troughs and the troughs crests. The node line is nearly normal to the line
joining the droplets, indicating that they are bouncing close to the angle with
the largest wave amplitude.  The pattern is not so evident in (a), due to the
greater angular velocity and the presence of higher-order components.

%

\subsection{Angular momentum}
\label{sec:irrotational-flow}

%

\begin{figure}[htb]
	\centering
		\includegraphics[width=0.45\textwidth]{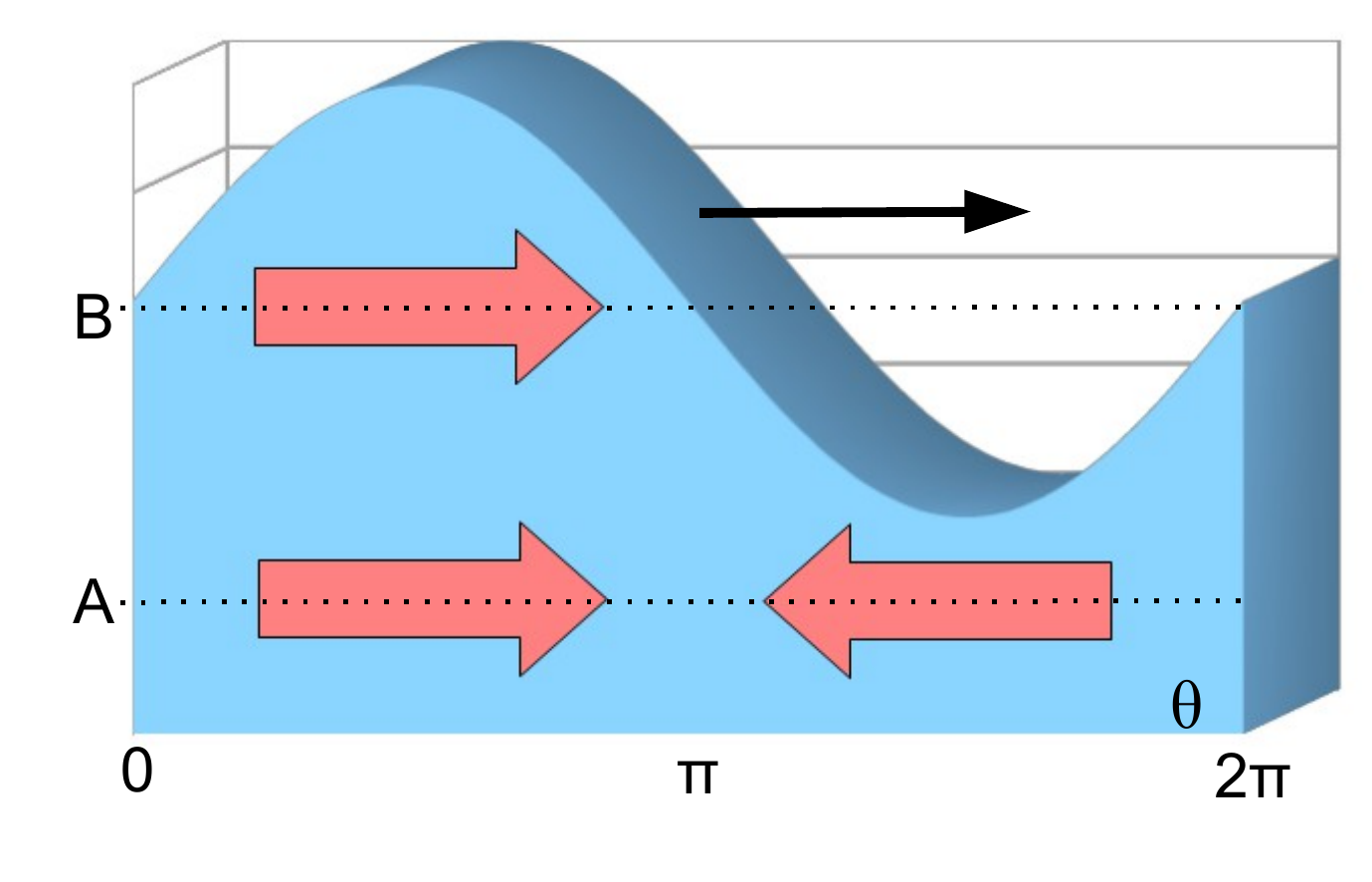}
	\caption{\em The wave $h_1$ in \eqref{eq:jm-cos} 
	at fixed radius at the instant $\omega_o t = \pi/2$.
	The flow speed (red arrows) is proportional to the wave height.}
	\label{fig:wave-angular}
\end{figure}

Figure \ref{fig:wave-angular} shows how the wave height $h_1$ in \eqref{eq:jm-cos}
varies with angle. The wave propagates in the $+\theta$ direction.
The flow velocity {\bf u} is irrotational ($\oint {\bf u.dl} = 0$) when 
the path of integration {\bf dl} is
on the submerged line $A$, but this does not mean the wave has no angular momentum
since it is not irrotational at $B$.
The elevations carry extra fluid around the centre. 

At large radius, the net flow around the centre 
approximates to that of a vortex when averaged over a period and a wavelength. 
The Bessel function in \eqref{eq:jm-cos}
approximates to a standing wave in the radial direction, whose amplitude
reduces as $A \sim r^{-\frac12}$. The flow speed is $u \propto A$ so the
net flow is proportional to $u A \sim r^{-1}$, which is the same as a
vortex. 

\begin{figure}[htb]
	\centering
		\includegraphics[width=0.45\textwidth]{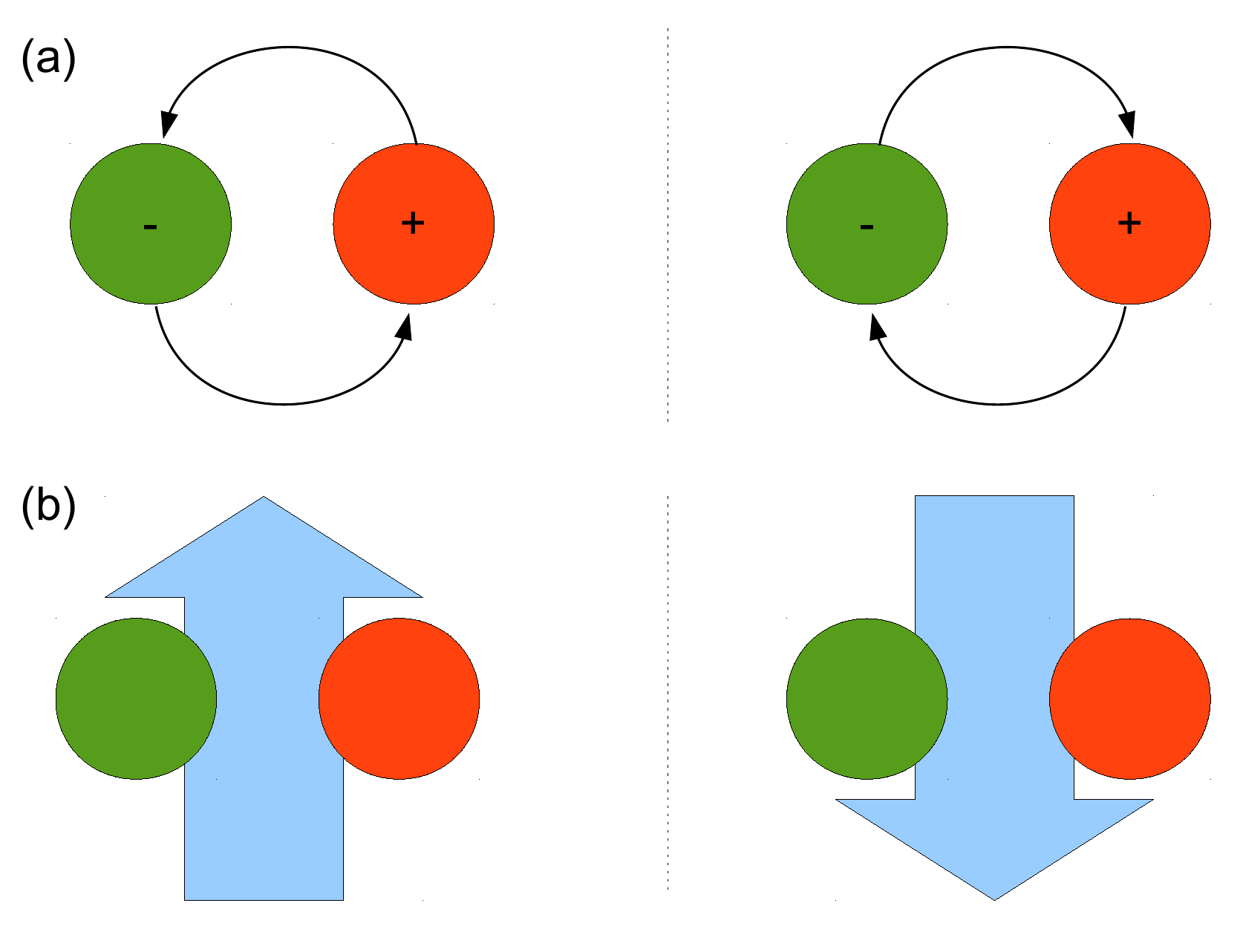}
                \caption{\em (a) Schematic drawing of the rotating droplet pair
                  photographed in figure \ref{fig:j1-droplets}(a), and its
                  image in the boundary.  At the instant drawn, the green
                  circles are the wave troughs (where the droplet lands) and
                  the red circles are crests.  (b) The fluid flow due to the
                  rotational motion.  The trough has negative volume and
                  contributes to the flow in the same Cartesian direction as
                  the crest.}
	\label{fig:droplet-pair-force}
\end{figure}

At small radius the flow diverges from that of a vortex, 
and in particular there is no singularity.
It is illustrated in figure \ref{fig:droplet-pair-force}.


\subsection{Attraction to the boundary}
\label{sec:eddy-attraction}

Two vortices of opposite circulations are
attracted towards one another because their flows reinforce in the region
between them, giving a reduced Bernoulli pressure. The rotating pairs should
similarly be attracted towards their images in the boundary, which rotate in
the opposite direction.  A protrusion on the boundary might be used to test for
this.

The fine structure constant of the interaction is as follows. We saw that
individual droplets are repelled from the boundary by an inverse square
force. This static force obeys the equations of electrostatics, with a fine
structure constant of $\alpha \sim 0.3$. We also saw evidence for a motion-dependent
force which obeys the same equations as magnetism, in which a droplet and its
image in the boundary are attracted towards one another when they both move in
the same direction parallel to the boundary.
	 
The static forces between a pair and its image in the boundary nearly cancel
out, since if one droplet of a pair is attracted to the image then the other,
being antiphase, will be repelled.  However, the motion-dependent force
is always attractive.  For example, consider the interactions with the wave
crest marked in red on the left of figure \ref{fig:droplet-pair-force}. It is
moving in the same direction as the (green) trough in the image, an alignment
which we saw produces a force of attraction
(section~\ref{sec:magnetic-interaction}).  The crest in the image (red) moves
in the opposite direction and with the opposite phase. Each of these
individually reverses the direction of the interaction, so the combination
leaves the sign unchanged.

We saw that the ratio of the moving to the static forces is $v^2/c^2$ (the same
as the ratio of the magnetic to the electrostatic forces for particles moving
at velocity $v$). The force on each droplet is doubled, since there are two
images, but it must be averaged over a rotation, giving a factor of $\frac12$.
The fine structure constant of the interaction is thus

\begin{equation}
	\alpha_2~~=~~ \frac{v^2}{c^2}\: \alpha_1
\label{eq:fine-struct-constant-pair}
\end{equation}

where $\alpha_1$ is the fine structure constant for the static force between
individual droplets. The strength of the interaction depends on the rotational
speed, which can be varied in the experiment.  A typical value might be $v =
\frac14 c$ and $\alpha_1 \sim 0.3$ giving $\alpha_2 \sim 1/50$.

Our model predicts that an orbiting droplet pair will be attracted
towards the boundary with this reduced fine structure constant. This phenomenon has been noted by the experimenters; orbiting pairs that approach a submerged boundary at a shallow angle can stick to it and then move along it, playing `hopscotch' as each droplet takes it in turn to leapfrog the one in front. However an experiment with precise measurements has not yet been performed. 

\subsection{The emergence of spin-half behaviour}
\label{sec:spin-half}

The rotating waves in \eqref{eq:jm-cos} can be treated as independent because they are 
orthogonal in the sense that
\begin{align}
	\int_0^{2 \pi} h_m ~h_n ~d\theta~~&=~~0&(m \neq n)
\label{eq:orthogonality-real}
\end{align}
as may be verified by direct substitution.

We have seen that the angular momentum of the wave $h_1$ in
\eqref{fig:j1-droplets} is in the $+z$ direction (vertically upwards), and it
is in the $-z$ direction for $h_{-1}$. The photograph in figure
\ref{fig:j1-droplets} shows the case where the waves have nearly equal
amplitude. What if the amplitudes are not equal?

It simplifies the analysis to consider `degenerate' solutions, that is,
solutions that have the same energy. (Solutions of arbitrary energy can be
obtained by scaling the wave height.) The degenerate solutions are
\begin{equation}
	h~~=~~\cos(\alpha) \: h_1 ~+~ \sin(\alpha) \: h_{-1}
\label{eq:spin-wave}
\end{equation}
where $\alpha$ is a real parameter. This is a solution to the wave equation
because it is a sum of solutions.  Its energy is proportional to $\cos^2 \alpha
+ \sin^2 \alpha$, which is constant, so the waves are degenerate. The angular
momentum is
\begin{align}
	L~~&=~~L_o (\cos^2 \alpha - \sin^2 \alpha) \nonumber
\\     &=~~L_o \cos(2 \alpha)
\label{eq:spin-angular-momentum}
\end{align}
where $L_o$ is the angular momentum of $h_1$. The angular momentum and the wave 
pattern $h$ vary continuously with the parameter $\alpha$, as shown in the
table below

\begin{center}
		\begin{tabular}{lll}
			$\alpha$ 
			& $L/L_o$
			& h
		\\ \hline \hline
			0 
			& 1
			& $h_1$ 
		\\ $\frac{\pi}{4} $
			& 0
			& $\frac{1}{\sqrt{2}}(h_{-1} + h_1)$
		\\ $\frac{\pi}{2} $
			& -1
			& $h_{-1}$
		\\ $\frac{3 \pi}{4} $
			& 0
			& $\frac{1}{\sqrt{2}}(h_{-1} - h_1)$
		\\  $\pi$ 
			& 1
			& $-h_1$ 
		\\ \hline
		
		\end{tabular}
\end{center}

As we can see in the table, the wave field reverses sign after the direction of the
angular momentum has gone through a complete cycle.  Two cycles are needed to
return to the starting position.

Fermions are like these waves, in that their wavefunctions reverse sign if the
direction of their angular momentum is rotated through $360^\circ$.  It is
commonly believed that this behaviour cannot emerge from classical
mechanics. However the rotating droplets show that this belief is wrong.

In fact, double symmetry is already known in systems that contain two harmonic sub-systems.
Leroy, Bacri, Hocquet and Devaud provided another example in 2006
when they showed that two weakly coupled pendula with nearly the same frequency
also have this symmetry~\cite{leroy2006simulating}.

\subsection{Bloch sphere}
\label{complex-plane-extension}

The elementary waves in \eqref{eq:jm-cos} are the real part of
\begin{equation}
	\xi_m~~=~~A ~e^{-i(\omega_o t - m \theta)} ~ J_m(k_r r)
\label{eq:jm-exp}
\end{equation}
where $A$ is the amplitude. This 
can be factored as before into
\begin{align}
	\xi~~&=~~\psi \: \chi \nonumber
\\	\psi~~&=~~e^{-i \omega_o t} \nonumber
\\	\chi~~&=~~A ~e^{i m \theta} ~ J_m(k_r r)
\label{eq:psi-chi-m}
\end{align}

The last section showed that $\psi$ obeys Schr\"odinger's equation. 
Now let us examine the factor $\chi$.

%


When the wave height in \eqref{eq:spin-wave} is extended into the complex plane 
as in \eqref{eq:psi-chi-m}, 
we get a simple way to provide an arbitrary origin of time
for each of the two components
\begin{equation}
	\chi=e^{iS} 
	\left[ 
			\cos \left( \tfrac12{\beta} \right) \chi_1
		+ e^{i \varphi} ~\sin \left( \tfrac12 \beta \right) \chi_{-1}
	\right] 
\label{eq:bloch-state}
\end{equation}
where $S$ is an arbitrary overall phase, $\varphi$ is the relative phase of 
the two components, and we have defined $\beta = 2 \alpha$.

\begin{figure}[htb]
	\centering
		\includegraphics[width=0.4\textwidth]{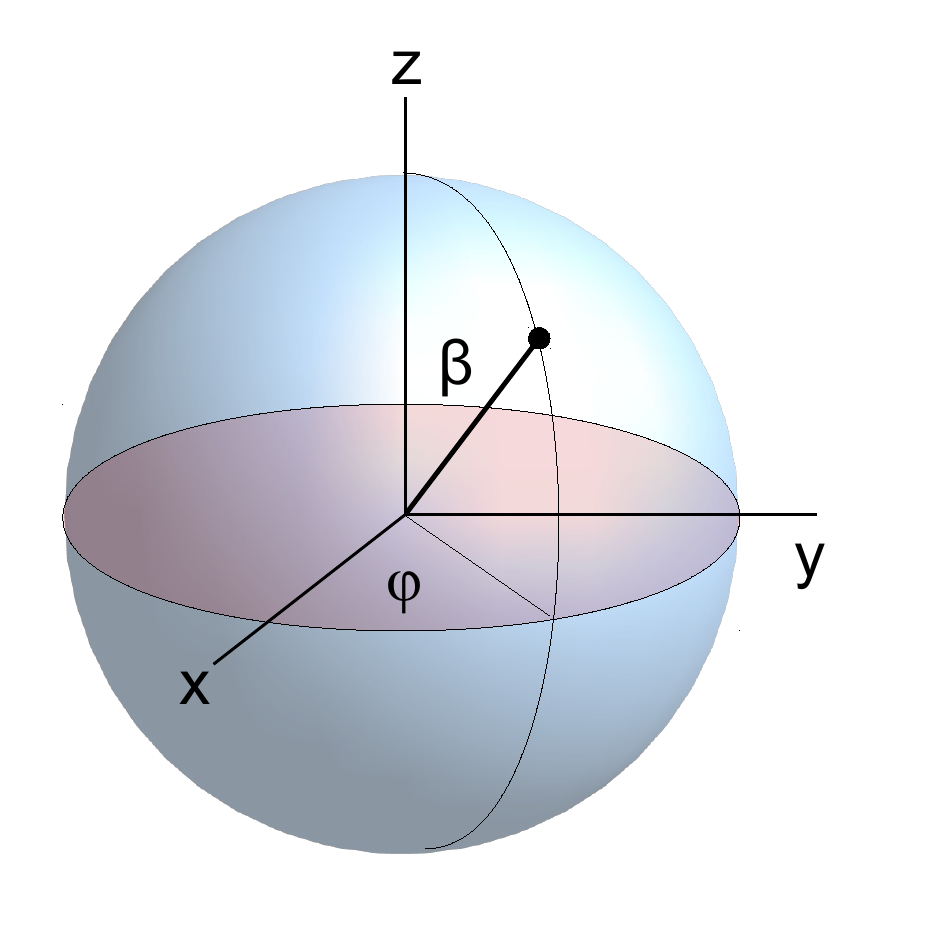}
                \caption{\em A Bloch sphere. }
	\label{fig:spherical-angles}
\end{figure}

The parameters in this equation can be represented on a sphere
as shown in figure \ref{fig:spherical-angles}. 
The angular momentum normal to the
surface (in the z direction) is proportional to $\cos \beta$.
When $\beta = \frac12 \pi$, the angular momentum vanishes and
there are standing waves whose amplitude is greatest at angle
$\varphi$ to the $x$ axis.
However, this diagram should not be over-interpreted.  It would be wrong to conclude
that the system is physically oriented in the direction indicated in the
figure. The existence of a simple geometrical way to picture the parameters 
in \eqref{eq:bloch-state}
should not blind us to the fact that we are describing
ordinary surface waves which cannot rotate out of the plane of the surface.

Nonetheless, equation \eqref{eq:bloch-state} is the same as 
that of a spin-half particle whose wavefunction is $\chi$ where
$\chi_1$ is the spin-up state and $\chi_{-1}$ the spin-down state, 
which is usually represented on the `Bloch sphere' in figure \ref{fig:spherical-angles}.
By inspection, the sign
of $\chi$ reverses when $\beta$ increases by $2 \pi$, which is characteristic of
spin-half systems.

\subsection{Pauli spin matrices}

The mapping of the wave height near a droplet onto the Bloch sphere can be shown more formally by 
writing \eqref{eq:bloch-state} as a dot product of two vectors
{\bf a} and $\boldsymbol\chi$
\[
	\chi ~~=~~ {\bf a}.\boldsymbol\chi~~=~~(a_1, a_2).(\chi_1, \chi_{-1})
\]
where the values of $a_i$ are obtained from \eqref{eq:bloch-state}.  The
angular momentum of the first component is proportional to $|a_1|^2$, and that of the second
component is proportional to $-|a_2|^2$, so the normalised total is
\begin{equation}
	\sigma_z 
	~~=~~ \frac{|a_1|^2 - |a_2|^2}{|a_1|^2 + |a_2|^2}
\label{eq:spinz}
\end{equation}

This can also be written
\begin{equation}
	\sigma_z ~~=~~ \frac{{\bf a}^* .~\widehat{\sigma}_z \: {\bf a}}
			   {{\bf a}^* . {\bf a}}
\label{eq:sigma-z}
\end{equation}
where $\widehat{\sigma}_z = (\begin{smallmatrix}
	1&0\\0&-1
	\end{smallmatrix}
	)$ is the same as the Pauli spin matrix for the $z$ direction.

As in quantum mechanics, we can extend this as follows. The Pauli matrices are 
\\
$\widehat{\sigma}_x = (
	\begin{smallmatrix}
	0&1\\1&0
	\end{smallmatrix}
	)$,~ 
$\widehat{\sigma}_y = (
	\begin{smallmatrix}
	0&-i\\i&0
	\end{smallmatrix}
	)$,~
$\widehat{\sigma}_z = (
	\begin{smallmatrix}
	1&0\\0&-1
	\end{smallmatrix}
	)$,
and spin projections $\sigma_i$ are defined by

\[
	\sigma_i ~~=~~ \frac{{\bf a}^* .~\widehat{\sigma}_i \: {\bf a}}
			   {{\bf a}^* . {\bf a}}
\]
where $i$ can be $x, y$ or $z$. The eigenvectors of $\widehat\sigma_i$ are

{\small
\begin{center}
		\begin{tabular}{lllcc}
				$\beta$ & $\varphi$ & $(a_1, a_2)$ & $(\sigma_x, \sigma_y, \sigma_z)$ & Eigenvector of
				\\ \hline \hline
				$\frac12 \pi$ & $0$ & $\frac1{\sqrt2} (1,1)$ & $(1,0,0)$ & $\widehat\sigma_x$
		\\		$\frac12 \pi$ & $\frac12 \pi$ & $\frac1{\sqrt2} (-i,i)$ & $(0,1,0)$ & $\widehat\sigma_y$
		\\		$0$ & $0$ & $(1,0)$ & $(0,0,1)$ & $\widehat\sigma_z$
		\\		\hline
		\end{tabular}
\end{center}
}

It will be noticed that $(\sigma_x, \sigma_y, \sigma_z)$ correspond to the
Cartesian coordinates of a unit vector at the spherical angle $(\beta,
\varphi)$ in figure \ref{fig:spherical-angles}. This is the basis of the Bloch
sphere, which maps between the two representations.  The mapping is a double
covering because $\chi$ reverses sign when $\beta$ increases by $2 \pi$.
The same mathematics is used to describe fermions in quantum mechanics.

%
%

\subsection{Antisymmetry}

\begin{figure}[htb]
	\centering
		\includegraphics[width=0.45\textwidth]{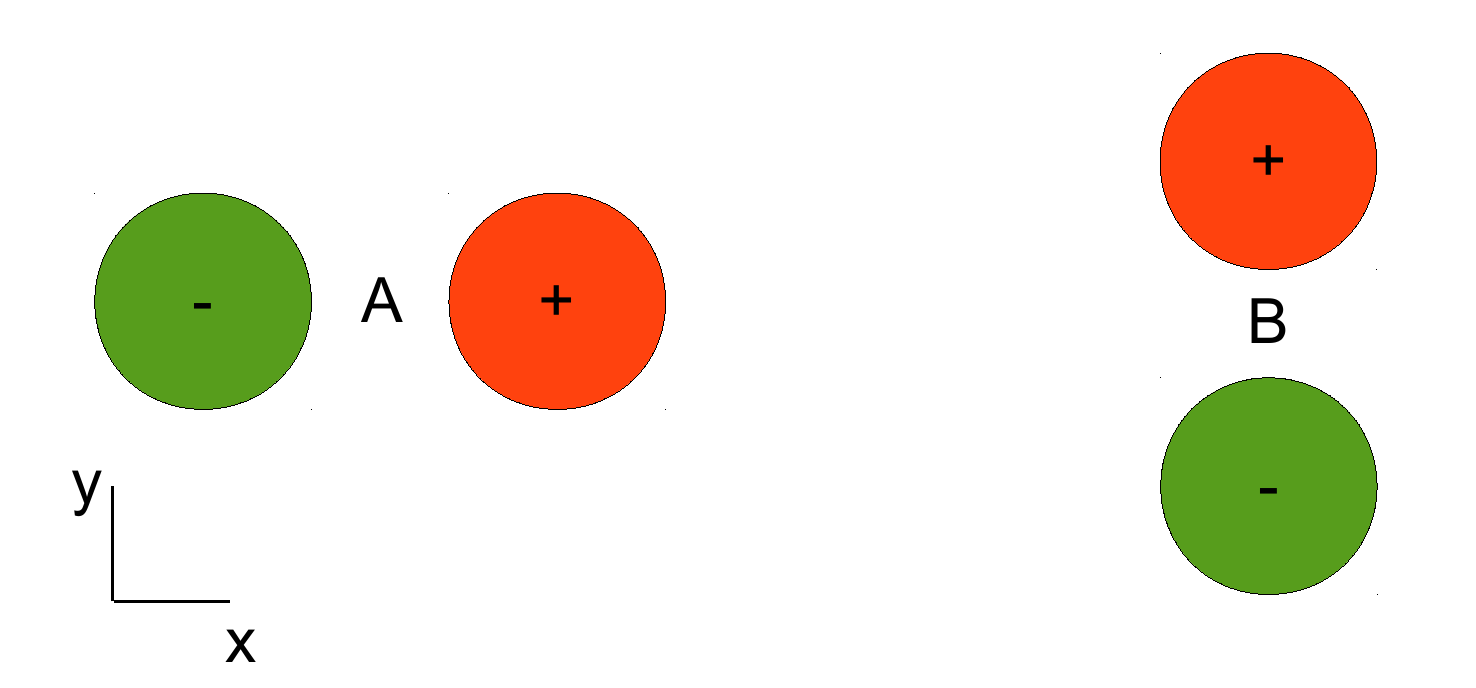}
                \caption{\em Schematic of two droplet pairs near each other. 
				Elevations are marked red and depressions green.}
	\label{fig:sigmax-y}
\end{figure}

When the driving amplitude is reduced, the rotation speed of the droplets photographed in figure
\ref{fig:j1-droplets} slows to zero.
Figure \ref{fig:sigmax-y} is a schematic of two droplet pairs near each other. 
$A$ is a solution to \eqref{eq:bloch-state} with $(\beta, \varphi) = (\frac12 \pi, 0)$ 
and $B$ has $(\frac12 \pi, \frac12 \pi)$.

The droplets in $B$ have opposite phases, so one will attract the other pair whilst the other
repels and the net forces cancel. However, there is still an effect involving orientation.
One droplet in $A$ is closer to $B$ than the other, which will cause $B$ to rotate
anticlockwise. This prediction might be tested experimentally.

After $B$ has rotated into the preferred alignment, the solutions will be oriented
in the $x$ direction and the wave height is the real part of
\begin{equation}
	\xi~~=~~\xi_a({\bf x}, t)  ~-~ \xi_a({\bf x - d}, t)
\label{eq:antisymmetric}
\end{equation}
where $\xi_a$ is the wave due to $A$ and $\bf d$ is the separation of the pair, $B - A$. 
 
Equation \eqref{eq:antisymmetric} is antisymmetric, and, in particular, exchanging $A$ and $B$ 
reverses the sign of the wave field.
This may be compared to the principle, formulated by Wolfgang Pauli in 1925,
that the total wave function for two identical fermions is anti-symmetric with respect to exchange of the particles.

\section{Discussion}

In this paper we have explained why the bouncing droplet experiments of Couder,
Fort and colleagues are a pretty good model for quantum mechanics. 

We have
derived from first principles that bouncing droplets are, to a rather good
approximation, Lorentz covariant, with $c$ being the speed of 
surface waves; that they obey an analogue of
Schr\"odinger's equation where Planck's constant is replaced by an appropriate
constant of the motion; that the force between them obeys Maxwell's equations, 
with an inverse-square attraction and
an analogue of the magnetic force; and finally that orbiting droplet pairs 
exhibit spin-half symmetry and align antisymmetrically as in the Pauli exclusion principle. 

These results explain why droplets undergo single-slit and double-slit
diffraction, tunnelling, Anderson localisation, and other behaviour normally associated with quantum
mechanical systems. We make testable predictions for the behaviour of droplets
near boundary intrusions, and for an analogue of polarised light.

The mathematical model described here may be useful as a teaching aid. 
Bouncing-droplet experiments have become popular with undergraduates; we show here that they 
can be explained with the mathematics routinely taught in a first undergraduate course in fluid 
mechanics. Indeed they are already sometimes one of the examples used to motivate such courses. 

For an introductory course in quantum mechanics, droplet models might help students overcome the initial feeling of bewilderment at the quantum-mechanical wavefunction $\psi$. Here, $\psi$ emerges naturally from known physical principles. This should help explain how the wavefunction of several interacting particles can emerge as a function of their position, momentum and spin, yet still be defined as a single amplitude and a single phase at each point in space -- helping students to avoid confusion over such concepts as configuration space and quantum entanglement. A vivid experimental model with a clear mathematical explanation should also help demystify spin-half behaviour and antisymmetry.

Finally, one might ask whether it is possible to extend this model from two dimensions to three. 
In separate
work we show a model of rotons in liquid helium with similar properties to the
droplets described here. Second sound in helium can be modelled as waves in a gas of
quasiparticles, the lambda point as its Kosterlitz-Thouless transition, while transverse sound 
emerges as the polarised-light analogue whose existence we predict here for droplet 
experiments~\cite{BradyAndersonbook}. 
So there may be room for further research on even more complex and realistic analogue models of quantum mechanics.

\section*{Acknowledgments}

We are very grateful to Yves Couder, Emmanuel Fort and Antonin Eddi not just for discussions but for access to their raw data and permission to use it here. We are also grateful to Robin Ball, John Bush, Graziano Brady, Basil Hiley, Keith Moffatt, Valeriy Sbitnev and to seminar participants at Warwick and Cambridge for comments, criticism and feedback.


\bibliography{droplets}
\bibliographystyle{unsrt}

\end{document}